\documentclass[12pt]{iopart}

\pdfoutput=1 
\usepackage{iopams} 
\usepackage{textcomp}
\usepackage{gensymb} 
\usepackage{graphicx}
\usepackage{subcaption}

\usepackage{hyperref}
\usepackage{enumitem}

\usepackage{booktabs}

\begin{document}
\bibliographystyle{vancouver}

\title[Deep learning based surrogate modeling for thermal plume prediction]{Deep learning based surrogate modeling for thermal plume prediction of groundwater heat pumps}
   

\author{Kyle Davis$^{1,*}$, Raphael Leiteritz$^{2,*}$, Dirk Pflüger$^{2}$, Miriam Schulte$^{1}$}
\address{1. Simulation of Large Systems, IPVS, University of Stuttgart, 70569, Stuttgart, Germany}
\address{2. Scientific Computing, IPVS, University of Stuttgart, 70569, Stuttgart, Germany}
\address{* Authors contributed equally.}
\ead{kyle.davis@ipvs.uni-stuttgart.de}
\ead{raphael.leiteritz@ipvs.uni-stuttgart.de}
\vspace{10pt}
\begin{indented}
\item[]February 2023
\end{indented}

\begin{abstract}
The ability for groundwater heat pumps to meet space heating and cooling demands without relying on fossil fuels, has
prompted their mass roll-out in dense urban environments. In regions with high subsurface groundwater flow rates, the
thermal plume generated from a heat pump’s injection well can propagate downstream, affecting surrounding users and
reducing their heat pump efficiency. To reduce the probability of interference, regulators often rely on simple analytical
models or high-fidelity groundwater simulations to determine the impact that a heat pump has on the subsurface aquifer
and surrounding heat pumps. These are either too inaccurate or too computationally expensive for everyday use.
In this work, a surrogate model was developed to provide a quick, high accuracy prediction tool of the thermal plume
generated by a heat pump within heterogeneous subsurface aquifers. Three variations of a convolutional neural network
were developed that accepts the known groundwater Darcy velocities as discrete 2D inputs and predicts the temperature
within the subsurface aquifer around the heat pump. A data set consisting of 800 numerical simulation samples, generated from random
permeability fields and pressure boundary conditions, was used to provide pseudo-randomized Darcy velocity fields as
input fields and the temperature field solution for training the network.
The subsurface temperature field output from
the network provides a more realistic temperature field that follows the Darcy velocity streamlines, while being orders
of magnitude faster than conventional high-fidelity solvers.
\end{abstract}

%
%
%
%
%

\section{Introduction}

Addressing the challenges of climate change and growing energy costs are forcing cities to increase their usage of renewable energies. 
The European Union 2030 climate and energy framework requires a minimum reduction of 40\% in greenhouse gas emissions, with a minimum of 32\% 
renewable energy usage \cite{eu2030report}.   
Energy usage can be significantly reduced by reducing the space heating and cooling demands of buildings, 
which have been met predominantly with fossil fuel based resources. 
An increasingly popular alternative is to use shallow groundwater heat pumps, which have proved to 
be an effective alternative within cities. 

Rolling out groundwater heat pumps on a city-wide scale is not without its challenges. 
City regulators must decide how many heat pumps can be effectively installed, how much can they be used and where can they be placed. 
The mass roll-out of heat pumps within a confined urban environment often leads to negative interference, where one heat pump changes 
the temperature of the water within a shared subsurface aquifer, reducing the efficiency of downstream heat pump systems. 
Additionally, the synergistic use of heat pump systems are often over-looked due to the added complexity of determining their mutual interactions. 
Therefore, monitoring any negative interactions and improving the overall 
efficiency by optimizing the usage and location of the heat pumps is imperative.

Numerous approaches to optimize heat pump usage and layout exist. 
Most methods involve either performing expensive numerical groundwater simulations using classical physics-based solver \cite{Attard2020,GarciaGil2020,Meng2019}, 
or by using a cheaper approximation of the heat transfer within the subsurface \cite{Beck2013,DePaly2012}. 
The classical numerical simulations offer highly accurate solutions for the subsurface thermal field, especially in the 
case of heterogeneous groundwater properties but at the expense of computational cost and increased run-time. 
The simulations often require extensive model preparation, calibration and validation, before running on large computing clusters. 
Only a few variations of the layout of heat pumps can be simulated within a reasonable time frame, limiting the 
practicality of this method for large cities with potentially thousands of heat pumps.  
Alternatively, using simplified approximations of the subsurface thermal profile allows for almost real-time solutions 
of the subsurface thermal profile induced by the heat pump, but at the expense of accuracy and only is valid for certain groundwater properties. 

A surrogate model that is able to reproduce the classical numerical simulation results, while being orders of magnitude cheaper, 
provides an attractive solution to the problem. 
This can be used to create a virtual model of the subsurface domain, allowing city and energy planners to perform optimization 
studies that are otherwise computationally infeasible, or provide an online monitoring tool of the subsurface of the city while 
having fast and accurate results.  
Many virtual models, or digital twins, of cities typically focus only on the above surface world and rarely model what happens underneath. 

In order to provide the level of accuracy as the high-fidelity solver, the surrogate model must be able to account for the subsurface heterogeneity. 
Popular methods for creating a reduced order model (ROM) are projection-based methods, such as proper 
orthogonal decomposition \cite{Cardoso2009,Pasetto2011,Li2013}.
Projection methods use snapshots of input-output pair information, where previous high-fidelity simulation runs are used to obtain these pairs. 
Once a ROM is generated, access to the underlying numerical model is typically required to build the surrogate model. 
This is sometimes problematic if the underlying numerical model is inaccessible, or if a truly black-box surrogate model is required. 

Recently, deep learning for black-box surrogate modeling of physical problems have been created using artificial 
neural networks (ANN) \cite{Vinuesa2022,Chen2021}, requiring access to the input-output data pairs only and not 
the underlying numerical model or internal simulation solver.  
ANNs have already been applied to modeling reacting flows \cite{Laubscher2017,Toit2018},  predicting airflow over 
airfoils \cite{Thuerey2019} and modeling fluid flow within porous media \cite{Wang2020,Tang2021}. 
In addition to surrogate modeling, ANNs can provide coarse to fine mesh mapping, where the network learns to approximate 
the fine scale information with access to the coarse scale features only \cite{Wang2021}, 
providing a middle-ground feature between a ROM and high-fidelity solution. 
Critical to the success of ANNs are the data, where generating sufficient simulation data to train an 
ANN is often a stumbling block for physics-based surrogate models.
Physics-informed neural networks have developed due to the underlying difficulty of obtaining expensive numerical simulation data \cite{Raissi2019}. 
Instead of using the input-output data pairs, the network is instead trained knowing that the 
output must satisfy a partial differential equation (PDE) that defines the physical problem. 
Therefore, only input data is required and the network loss function consists of the mismatch 
between the output result and the PDE that it must satisfy. 
This has been applied to numerous physics-based surrogate models where the governing equations are known \cite{Zhu2019,Kashefi2021,Laubscher2021}. 

The objective of our study was to utilize artificial neural networks to create a fast surrogate model that 
solves for the local subsurface thermal field due to the presence of a groundwater heat pump, while 
accounting for the subsurface heterogeneity of the permeability and Darcy velocity fields. 
The surrogate model can be used to perform heat pump layout and usage optimization, which may consist 
of thousands of layout configurations, or serve as a fast approximation to build into an online evaluation tool for groundwater heat pump management.

\section{Method}

\subsection{Open-loop groundwater heat pumps}

\begin{figure}
\centering
\includegraphics[width=0.5\textwidth]{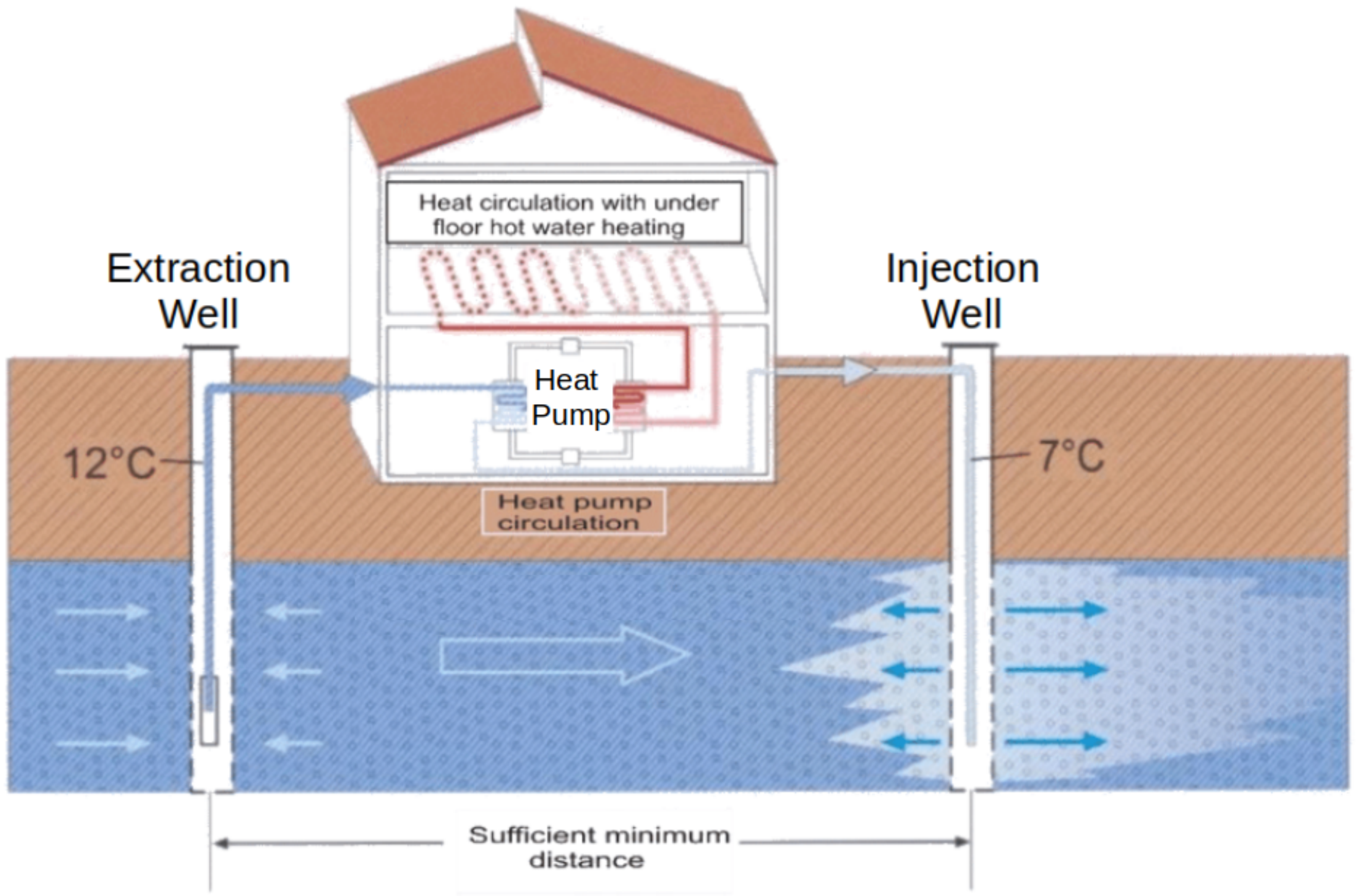}
\caption{\small{Open loop groundwater heat pump, where the flow of groundwater causes a plume to develop that stretches downstream of the injection well. Image modified from \cite{Pavlova2016}}}
\label{fig:gwhp}
\end{figure}

Shallow groundwater heat pumps are devices that transfer heat to and from a shallow aquifer beneath the surface in order to provide space heating or cooling. 
Open-loop groundwater heat pumps, depicted in figure \ref{fig:gwhp}, function by extracting water 
from the subsurface aquifer, passes the water through a heat exchanger, before re-injecting the water back into the subsurface. 
When operating in heating mode, i.e., heating a building, the energy in the water is passed 
into the building through the heat exchanger, thereby cooling the water and re-injecting the 
fluid at a lower temperature back into the aquifer. 
Conversely, in cooling mode the energy is passed from the building and into the fluid, warming 
up the water and re-injecting the warmer fluid back into the aquifer. 
The natural flow of the groundwater within the subsurface aquifer causes the re-injected water, 
now at an elevated or reduced temperature to the rest of the aquifer, to be pulled along with the movement 
of the groundwater, causing a thermal plume to form downstream of the injection well. 
If this plume reaches the extraction well of a nearby downstream heat pump, the heightened or 
lowered temperature of the groundwater within the plume may reduce the efficiency of the downstream system to the extent that it becomes unusable.  
When a small urban area is densely populated with heat pumps, the overall efficiency 
may be reduced more than if only a handful of well-placed heat pumps were used. 

\subsection{Subsurface modelling}

\begin{figure}
    \centering
    \begin{subfigure}[b]{0.4\linewidth}
    \includegraphics[width=0.8\textwidth]{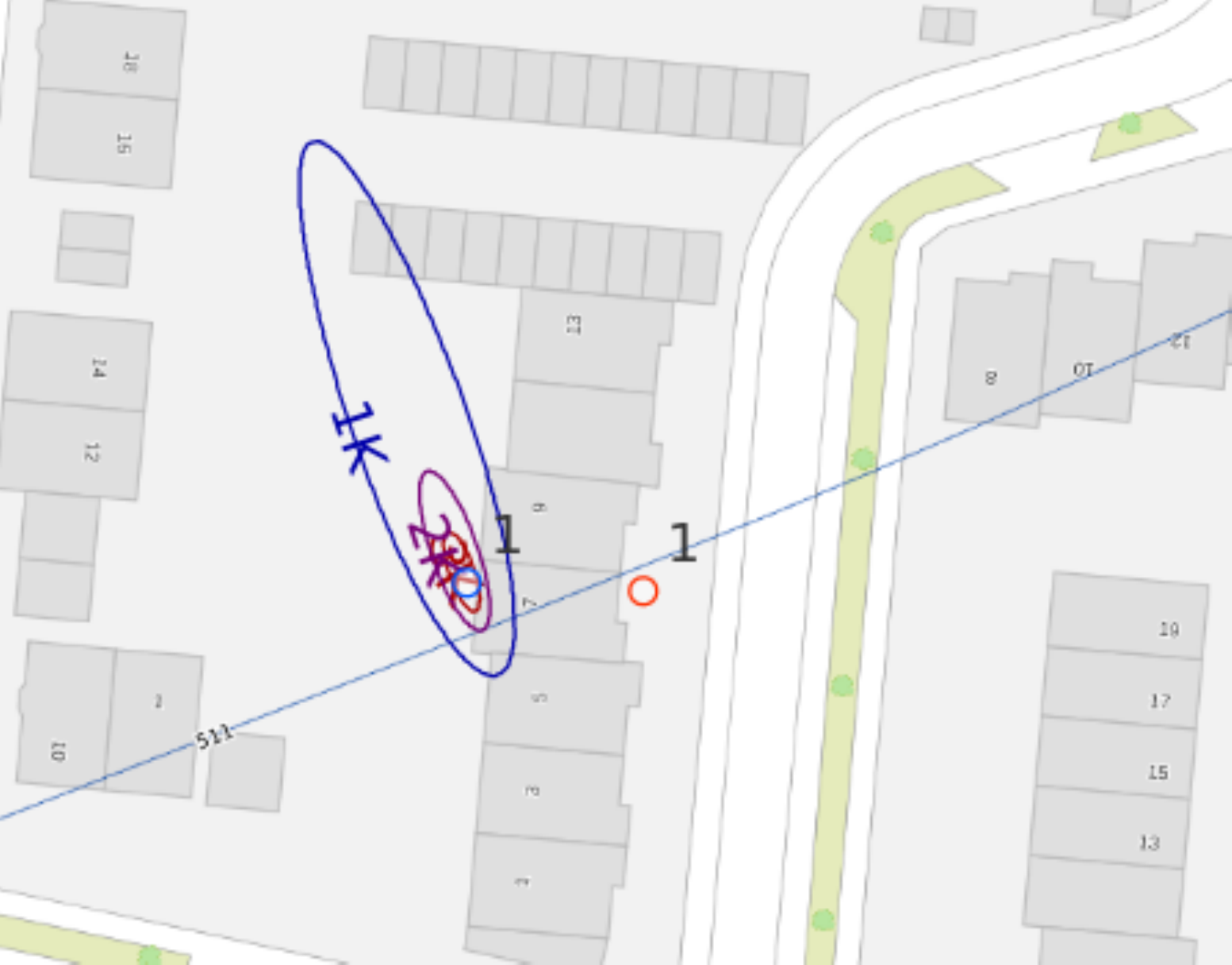}
    \caption{\small{Online evaluation tool}}
    \end{subfigure}%
    \hspace{1cm}
    \begin{subfigure}[b]{0.4\linewidth}
    \includegraphics[width=0.8\textwidth]{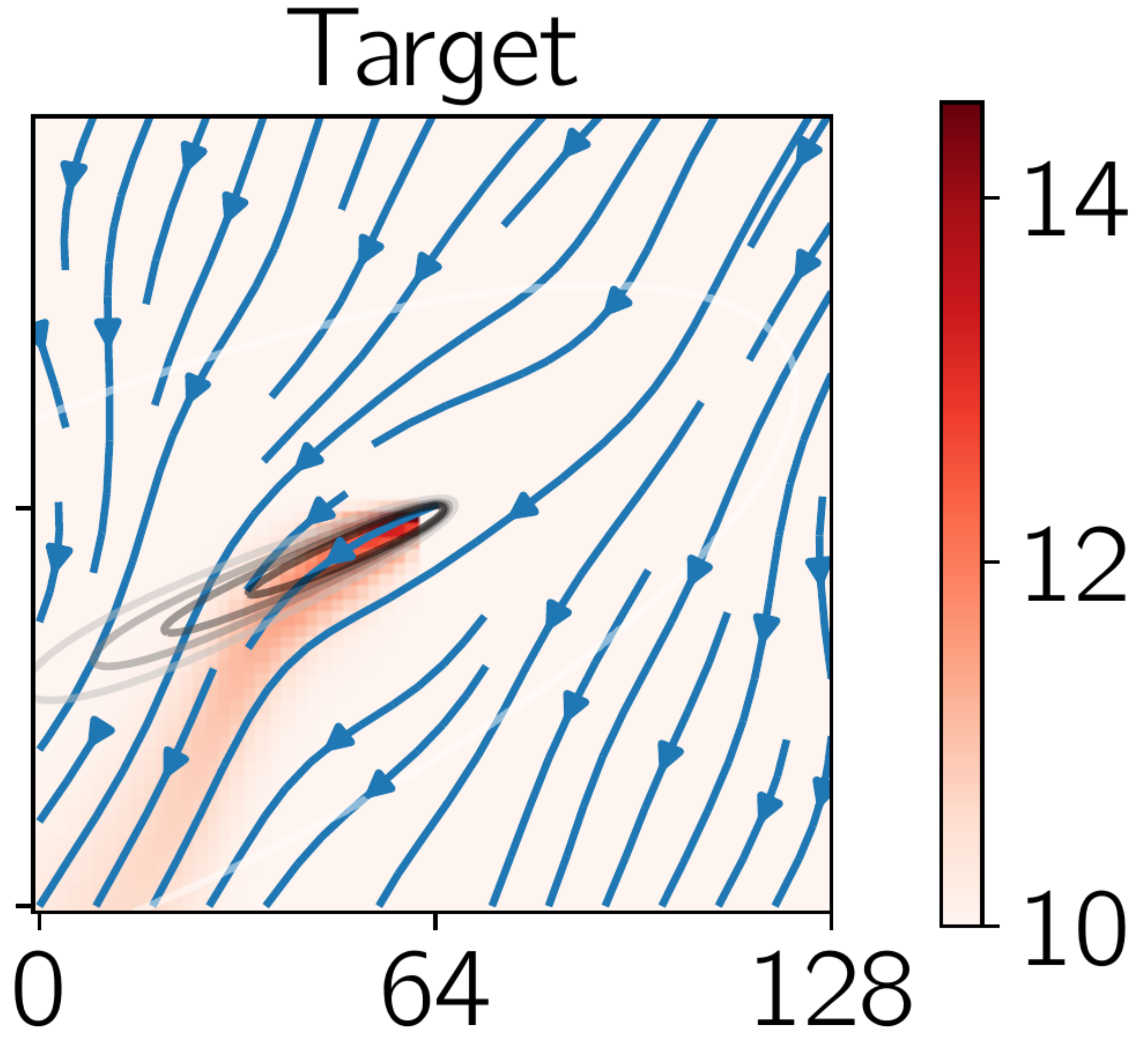}
    \caption{\small{Analytical vs. numerical solution}}
    \end{subfigure}%
    \caption{\small{\label{fig:ana_comp}Temperature iso-lines of the LAHM analytical solution applied to an online groundwater heat pump planning and optimization tool (left) and a comparison between the LAHM analytical solution temperature iso-lines overlaid onto the 2D numerical solution (right).}}
\end{figure}

Before any heat pump layout optimization study or heat pump monitoring can be performed, a suitable model for the groundwater temperature is required. 
There are two popular methods that exist to model the thermal plume that develops from an open-loop heat pump. 
The linear advective heat transport model (LAHM) from \cite{Kinzelbach1987} is a common analytical 
formulation that defines the change in groundwater temperature around the heat pump over time compared to the far-away background temperature 

\begin{eqnarray}
    \label{eqn:lahm}
    \Delta T (x, y, t) = \frac{Q \cdot \Delta T_{inj}}{4 \cdot n_e \cdot M \cdot v_a \cdot \sqrt{\pi \cdot \alpha_T}} \cdot exp \left( \frac{x - r}{2 \cdot \alpha_L} \right) \cdot \nonumber \\
    \frac{1}{\sqrt{r}} \cdot erfc \left( \frac{r - v_a \cdot t/R}{2 \cdot \sqrt{v_a \cdot \alpha_L \cdot t/R}} \right),
\end{eqnarray}

where $\Delta T (x, y, t)$ is the time-dependent temperature difference at coordinates $x$ and $y$ (the heat pump 
is at coordinates $x$ = 0 and $y$ = 0) at time $t$, $Q$ is the injection mass flow rate, $\Delta T_{inj}$ is the 
difference between the background temperature and the injection temperature, $v_a$ is the groundwater velocity at 
the heat pump injection well, $r$ is the radial distance from the injection well, $M$ is the aquifer thickness, 
and $\alpha_L$ and $\alpha_T$ are the longitudinal and tangential dispersivity values. 
The disadvantage of the LAHM is that the plume is dependent only on the velocity magnitude in subsurface aquifer 
at the heat pump location with constant subsurface properties. 
Therefore, no heterogeneity of the subsurface is accommodated. 
The uni-directional thermal plumes determined using the LAHM are shown in figure \ref{fig:ana_comp} as examples of the solution.  

The LAHM has already been integrated into an online evaluation tool shown in figure \ref{fig:ana_comp} (a), 
that determines the iso-lines of the change in subsurface temperature $\Delta T (x, y, t)$ due to the presence of a heat pump. 
This provides a fast surrogate model, but does not model the heterogeneity as shown in figure \ref{fig:ana_comp} (b), 
where the LAHM solution iso-lines are overlaid onto the classical physics numerical simulation results using a 3D groundwater 
mechanics solver. The classical solver accounts for subsurface heterogeneity, 
causing non-uniform Darcy velocity streamlines and varying direction of the thermal plume.  

More complex 3D numerical groundwater models can provide a more accurate solution for the groundwater temperature field. 
The governing equations for the subsurface groundwater model is 
derived from \cite{pflotrantheory}. 
The subsurface model applies to a single phase and variably saturated fluid, defined on a single space-time domain. 
The subsurface fluid flow is governed first by the conservation of mass,

\begin{equation}
\label{eqn:richards}
\frac{\delta}{\delta t} \left( \varphi s \eta \right) + \nabla \cdot \left( \eta \mathbf{q} \right) = Q_w , 
\end{equation}

with the porosity $\varphi$ [-], saturation ratio $s$ $[m^3 m^{-3}]$, molar density $\eta$ $[kmol m^{-3}]$, 
the Darcy velocity $\mathbf{q}$ and mass source/sink term $Q_w$ $[kmol \cdot m^{-3} \cdot s^{-1}]$. 
The groundwater temperature is modeled by the inclusion of the conservation of energy, 
\begin{eqnarray}
\label{eqn:thermo_equation}
\frac{\delta}{\delta t} \left( \varphi s \eta U + \left( 1 - \varphi \right) \rho_p c_p T \right) \nonumber \\
 + \nabla \cdot \left( \eta \mathbf{q} H - \kappa \nabla T \right) = Q_e ,
\end{eqnarray}

with the rock density $\rho _r$, heat capacity $c_p$ and thermal conductivity $\kappa$ of the porous 
medium - fluid mixture and the energy source/sink term $Q_e$. The fluid enthalpy $H$ is related to the internal energy of the water through the expression 
\begin{equation}
\label{eqn:th_mode_2}
U = H - \frac{P}{\eta}. 
\end{equation} 

The Darcy velocity $\mathbf{q} = (q_x , q_y , q_z)^T$ in $[m \cdot s^{-1}]$ is defined as
\begin{equation}
\label{eqn:darcy_flux}
\mathbf{q} = - \frac{\mathbf{K}(s)}{\mu} \nabla \left( P - \rho g z \right) ,
\end{equation}

with the relative permeability field $\mathbf{K}(s)$ $[m^2]$, the viscosity $\mu$ $[Pa \cdot s]$, 
subsurface water pressure $P$ $[Pa]$, gravitational constant $g$ $[m \cdot s^{-2}]$ and the relative reference height $z$ $[m]$. 
The constant pressure initial condition for the subsurface model is defined on the boundary of the domain $\delta \Omega_{P}$, 
inducing the Darcy velocity $\mathbf{q} =  \mathbf{K} \nabla P$ throughout $\Omega_{P}$.
The boundary condition of a heat pump can be specified by either injecting energy into the domain 
via the heat flux $Q_e$, or by injecting a mass of water $Q_w$ at a predetermined temperature. 
The latter method more resembles the operation of an open-loop heat pump, which pumps water back 
into the subsurface at a new temperature that is dependent on the extraction well temperature.  

The operation of commercial heat pumps utilize a constant temperature difference across the heat 
exchanger, and alters the amount of energy passed to/from the groundwater by increasing or 
decreasing the mass flow rate extracted from the subsurface aquifer. 
The difference between the injection well $T_{inj}$ and extraction well $T_{ext}$ temperature is 
defined as $\Delta T = T_{inj} - T_{ext} = 5$ when operation in cooling mode and $\Delta T = -5$ when 
operating in heating mode. The extraction and injection mass flow rate is defined 
from the amount of energy passed into-or-from the fluid and $\Delta T$ by 
\begin{equation}
\dot{m} =  \frac{\dot{Q}}{c_p \Delta T} \, ,
\label{eqn:hp_formula}
\end{equation}
with the energy transferred to/from the fluid $\dot{Q}$ $[J]$, the specific heat of 
water (assumed constant) $c_p$ = 4184 $[J \cdot kg^{-1} \cdot K^{-1}]$ and the temperature difference $\Delta T$. 

\subsection{Neural network design}

\subsubsection{Input data:} 

The neural network operation is divided into two parts: offline and online stages. 
The offline stage involves generating and processing the training data, training the network and finally testing the network accuracy. 
The online stage involves feeding the input to the network and obtaining the output during normal usage of the network.  
The offline step is typically computationally expensive, but is tolerable as this is only performed once in order to generate a 
network. 
The online step is intended to be performed thousands if not millions of times, thereby requiring a fast prediction performance in the online step. 
To minimize the computational cost of the network prediction step, the surrogate model must not 
rely on performing any numerical simulation during the online step and therefore, all input data must be readily available. 
A baseline simulation of the entire subsurface of Munich already exists and contains the subsurface 
permeability, pressure and Darcy velocity throughout the city's subsurface.
These data can simply be extracted from the baseline simulation result and input into the network 
during the online phase without having to perform any expensive numerical simulation. 

For this study of building a surrogate model, we ignore the time-dependent behavior 
of the thermal plume propagation and only consider the steady state solution. 
Examining equation \ref{eqn:thermo_equation} and ignoring the time-derivative, the thermal plume is dependent on an advection term and a diffusion term. 
We assume that the plume behavior is dominated by the advection term of this equation, where figure \ref{fig:ana_comp} 
strengthens this assumption as the thermal plume tends to follow the velocity streamlines. 
As the subsurface Darcy velocity is readily available and likely to provide meaningful information to the 
network in order to predict the plume behavior, we use the Darcy velocity components $q_x$ and $q_y$ as inputs into the network. 
Previous work on building a surrogate model of the subsurface flow accepted the groundwater permeability 
as input and predicted the groundwater pressure and Darcy velocities \cite{Zhu2019}. 
However, this is only applicable for scenarios with a constant pressure boundary condition, and therefore 
not applicable for general use such as with our surrogate model. 

\subsubsection{Output data:}

Obtaining the output results from the surrogate model must be orders of magnitude faster to 
attain than the high-fidelity model in order to be useful, while remaining accurate. 
For our specific application of optimization and condition monitoring, only the temperature 
output is required for build a practical surrogate model. 
The network input data (obtained from the baseline simulation data), does not include the influence 
that the heat pump has on the Darcy velocities or thermal field. 
The baseline simulations do not have the required output data necessary to train the network using supervised learning, which must be sourced elsewhere. 
Therefore, smaller 2D numerical simulations are performed in order to generate input-output data pairs for the network training and testing. 
However, these smaller numerical simulations are much smaller than the baseline simulation case, such that the data generation time is feasible for our model. 

\subsubsection{Network architecture:}

The available input and output data structure must be considered when designing the network architecture.  
The output of the 2D numerical groundwater simulation, which will be used as input and output of the network, 
can be treated as an image where the value at each pixel defines the Darcy velocity (input) or temperature (output) at the cell centers of the finite volume mesh. 
The input data are provided as a two-channel image $\varphi _{in} \in \mathbb{R}^{2\times65\times65}$ of the 
Darcy velocity $q_x$ and $q_y$ of size 65$\times$65 pixels each ($\varphi _{in,1}$ and $\varphi _{in,2}$). 
The network output $\varphi _{out} \in \mathbb{R}^{65\times65}$ is a single channel image of 65$\times$65 of the temperature field. 
The location of the heat pump is always at the center pixel. 
An image size of 65$\times$65 was selected so that an equal number of pixels surround the center pixel in each direction. 
The image-like nature of the input and output data naturally allows the use of a convolutional neural network (CNN), which is favorable for image like data. 
Previous work on predicting the behavior of physical systems have shown that CNNs are favorable using image-like data \cite{Thuerey2019,Zhu2019,Wang2021,Tang2020}. 

Within this work, we built and evaluated three variations of the TurbNet architecture 
by Thuerney et. al. \cite{Thuerey2019}, which is also a variant of the U-Net architecture by Ronneberger et. al. \cite{Ronneberger2015}. 
The TurbNet architecture features skip connections between the encoding and decoding steps.

Each layer of the CNN applies a 2D convolution operation to the input image by sweeping a 
fixed size kernel (e.g. 4x4) over the input and multiplying it with the underlying image data.
The kernel weights $\mathbf{W}_c$, are parameters which are optimized during the network's training procedure.
By having these weights shared for the whole image, the convolutional layer essentially learns to pick up translation invariant features.
A user-defined number of features, each having its own kernel, can then be learned to create multiple outputs per layer.
Usually, after each convolutional layer an aggregation operation is applied such as a 
pooling operation where multiple pixels are combined to create a coarser output.
Recursively continuing this process allows the network to pick up different features at multiple scales.

For the U-Net architecture this process is reversed after reaching a pre-defined bottleneck state where the coarsest features are represented.
This bottleneck is then followed by inverse convolutions to reconstruct higher-resolution images for the output path.
This is comparable to simple auto-encoder architectures which also feature an encoder-bottleneck-decoder structure.

The entire encoding-decoding procedure can be characterized as a mapping $S$, from the input images and kernel weights to the prediction output $\mathbf{\varphi}_{pred}$,

\begin{equation}
    S \left( \mathbf{W}_{c}, \mathbf{\varphi}_{in,1}, \mathbf{\varphi}_{in,2} \right) \mapsto \mathbf{\varphi}_{pred}.
\end{equation}
The network is trained to minimize the difference between the known output $\mathbf{\varphi}_{out}$ and the network 
prediction output $\mathbf{\varphi}_{pred}$ by varying the kernel weights $\mathbf{W}_c$ and other network parameters $\theta$


\begin{equation}
    \mathop{\mathrm{arg\,min}}_{W_c,\theta} \sum_{i}^{N_{data}} \Vert \mathbf{\varphi}_{pred} - \mathbf{\varphi}_{out} \Vert .
\end{equation}

Both the encoding and decoding steps are divided into multiple layers, which each contain a 
rectified linear unit (ReLU) activation function with a max pooling layer of 2$\times$2, a 
stride of $2$ for down-sampling the image size and batch normalization. The image down-sampling 
halves the image pixel size per direction (the first convolution layer only reduces the image size from 65$\times$65 to 64$\times$64). 
The total number of trainable parameters of the network can be increased by increasing the number 
of layers or increasing the number of initial features. This also affects the accuracy of the network. 

\begin{figure}
\centering
    \includegraphics[width=0.8\textwidth]{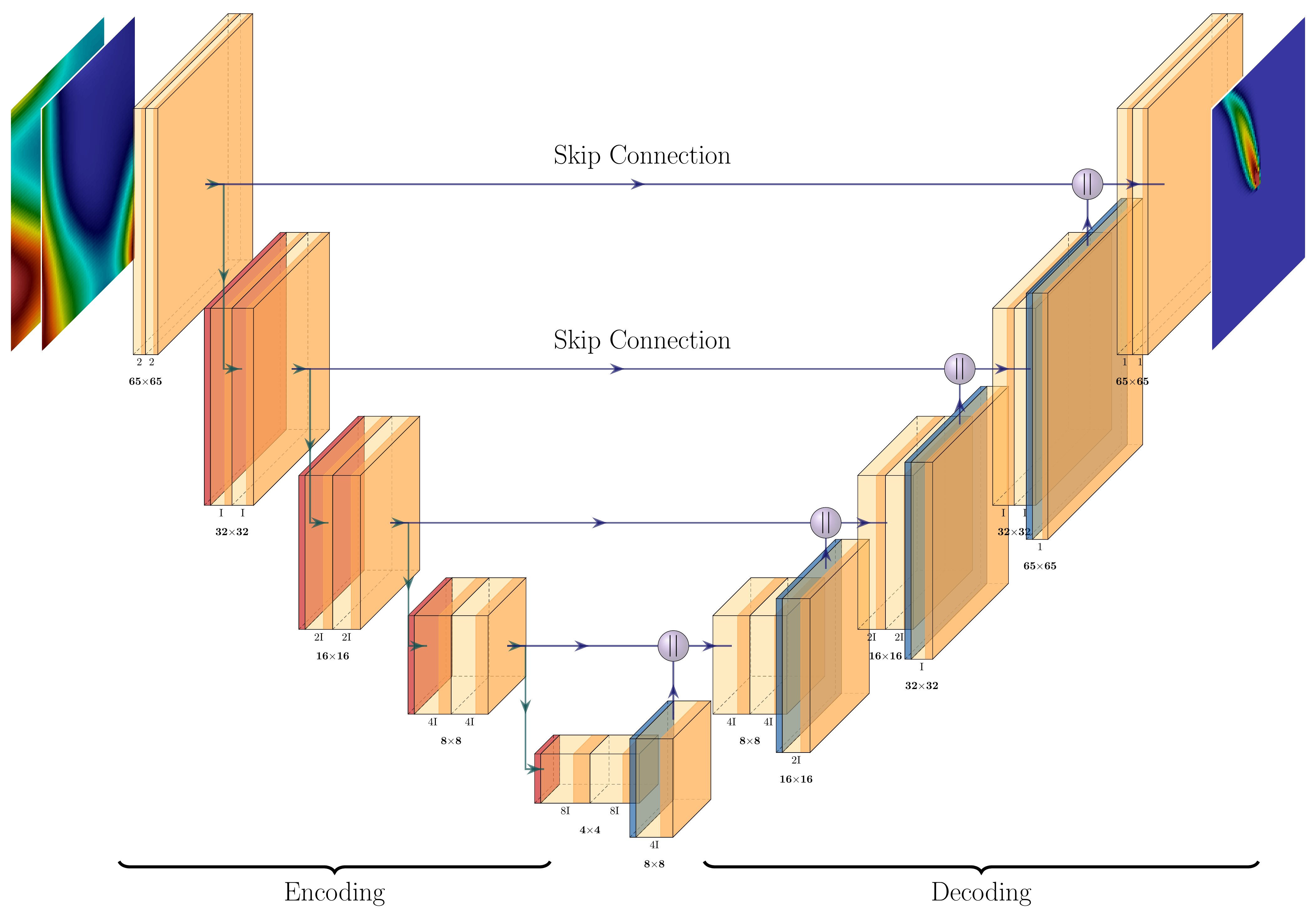}
    \caption[ ]{\label{fig:arch} \small{TurbNet-Geo architecture containing skip connections for the temperature plume prediction. The network accepts a 2-channel input of the $x$ and $y$-direction Darcy velocity magnitudes and outputs a single channel of the temperature field. The bottleneck contains 32$\times$ the amount of initial features. 
    }}
\end{figure}

The three architectures tested are:
\begin{enumerate}
\item TurbNet-Geo: 6 layers with skip connections
\item TurbNet-Geo-Light: 4 layers with skip connections.
\item TurbNet-Geo-NoSkip-Light: 4 layers without skip connections (TurbNet-Geo-Light without the connections).
\end{enumerate}
With the latter two architectures, we test the model performance with regard to fewer network parameters and the elimination of skip connections. 
The skip connections add a concatenation step between the encoding and decoding layers in order to retain features in the encoding steps. 
The TurbNet-Geo-Light network architecture is illustrated in figure~\ref{fig:arch}. 
The 2 channel input is provided on the left, with the network output on the right. 

\subsection{Data generation}

In order to study whether the CNN is able to predict the thermal plume with sufficient accuracy, 
a simple 2D groundwater model is utilized. This is sufficient for two reasons: firstly, the network 
only accepts 2D images as inputs and secondly, the thermal plume only influences the local area around a heat pump.  
This allows for relatively cheap numerical simulations to be performed to obtain the Darcy velocities 
without the influence of the heat pump, followed by re-running the same model but with the heat pump 
activated to obtain the thermal plume profile. 
As the thermal plume profiles are not available from the current baseline simulation, it would be 
too costly to run multiple large, high-fidelity simulations of a large area of the city, only to obtain local temperature profiles around the heat pumps. 
Therefore, the data generation procedure is divided into 4 steps:
\begin{enumerate}
    \item Input field generation - generates the numerical simulation input files with randomized permeability fields and pressure boundary conditions,
    \item Without-heat-pump evaluation - generates the Darcy velocity fields without the influence of a heat pump,
    \item With-heat-pump evaluation - the "Without-heat-pump" simulation is repeated but with the heat pump activated,
    \item Data manipulation - pre-processing of the input and output data for the network.
\end{enumerate}

All input and output training and testing data of the numerical groundwater simulations were generated 
using the subsurface simulation software PFLOTRAN v3 \cite{pflotran-paper}. The data generation simulations were performed on a workstation with an AMD Ryzen Threadripper 3960X 24-core Processor at the University of Stuttgart, running Ubuntu 20.04.

\subsubsection{Input field generation:}

The prediction quality of the network is dependent on the quality of the training data.
Therefore, the purpose of the input field generation procedure is to generate a variety of Darcy velocities and flow directions.
To generate the variety of velocity data, randomized permeability fields are combined with randomized pressure gradient boundary conditions to generate seemingly arbitrary Darcy velocities that are within a suitable velocity range.
For each data sample (one permeability field and one pressure gradient boundary condition), two simulations are performed: one with no heat pump active and, therefore, generating a velocity field without a heat pump. The second simulation has an active heat pump, injecting water at a different temperature than the background groundwater temperature to generate the thermal plume. 
 
The simulation domain covers an area of 130m$\times$130m$\times$2m, divided into 65$\times$65$\times$1 finite volume grid cells (each grid cell is 2m$\times$2m$\times$2m) for a total of 4225 cells.
The heat pump is placed at location $(33,33)$ to be in the center of the domain.
A permeability field is generated by placing pilot points on a uniformly space grid of either 4$\times$4 (16 pilot points) or 6$\times$6 (36 pilot points) within the simulation domain.
A random value is generated for each pilot point between 1.13 $\times 10^{-7}$ and 3.77 $\times 10^{-11}$.
The values at the pilot points are interpolated onto the PFLOTRAN mesh using radial basis function interpolation with thin-plate-splines basis function, to create the randomized permeability field.
The interpolation step from a coarse mesh of pilot points to a fine PFLOTRAN mesh avoids sudden changes in the permeability, which could occur if a random value was assigned to each mesh cell.  
The large difference in orders of magnitudes creates both high and low permeability regions, that can allow for the non-uniform velocity streamlines.  
The pressure gradient boundary condition is generated by randomly selecting two values between $[-0.0006,0.0006]$ and applying these in the $x$-direction and $y$-direction.
The magnitude of the values generates realistic groundwater Darcy velocities and allow the direction of flow at the heat pump location to be in any 360$\degree$ direction. 

Finally, the combination of each permeability field and pressure boundary condition resulted in 800 unique input data samples. The small PFLOTRAN simulation domain allowed for the fast generation of data, with each set of 25 samples taking approximately $180s$. 

\subsubsection{Without-heat-pump evaluation:}

The input data for the CNN requires the Darcy velocities without the influence of a heat pump in the region.
This recreates the conditions of extracting the velocities from a region of the baseline simulation domain without a heat pump.
Therefore, the heat pump mass flow rate was set to zero and the PFLOTRAN simulation was run for a period of 365 days, where the Darcy velocities were extracted at day 365 to obtain a pseudo steady-state solution. 

\subsubsection{With-heat-pump evaluation:}

Supervised training of the network requires the correct temperature field to evaluate the prediction accuracy.
Therefore, each simulation performed in the "Without-heat-pump evaluation" was rerun with the heat pump activated.
The heat pump mass flow rate was set to 0.05 $l/s$ and an injection temperature of 15$\degree$C, against the background temperature of 10$\degree$C. 

\subsubsection{Data pre-processing:}

The accuracy of the CNN model can be greatly improved by suitable pre-processing of the data. 
Firstly, the background of 10$\degree$C is subtracted from the temperature output $T_{Target}$ such that the far away temperature is around 0$\degree$C.
Next, each data sample field (Darcy velocities $q_x$ and $q_y$ and temperature field $T_{Target}$) is normalized to the range $[-1,1]$ over the whole data set.
To obtain the correct output, the inverse operation is applied to the temperature prediction to obtain the final output. 

\subsubsection{Network limitations:}

The current network design still suffers from some practical limitations. 
Firstly, the network can only predict the groundwater temperature under the assumption that the background temperature is completely uniform. 
Secondly, it is assumed that the injection temperature is exactly 5$\degree$C higher than the background temperature and with a constant flow rate of 0.05 l/s. 
However, the purpose of this study was to determine whether CNNs are capable of being a reasonably good surrogate model. 
Tackling further practical usage aspects are planned for future studies. 

\subsection{Experimental setup}

The three networks were trained using the Adam optimizer \cite{Kingma2014} for 50,000 epochs with a fixed learning rate of $0.0005$ and a batch size of $64$. 
Each network was trained using $4$, $8$, $16$ and $32$ initial features, with the total number of trainable parameters per network shown in table \ref{tab:net_arch_size}. 

\begin{table}[!hb]
  \centering
  \caption{\small{Number of trainable parameters for the convolutional neural network. The number of parameters increases when increasing the number of initial features extracted in the first convolutional layer.}}
  \label{tab:net_arch_size}
    \small{
\begin{tabular}{lcccc}
\toprule
Init. Feat. &  & TurbNetGeo         & TurbNetGeo-Light    &  TurbNetGeo-NoSkip-Light                  \\
\midrule
4  &          &          77,033            &         18,697            &       17,221             \\
8  &          &          306,257            &         73,873           &         68,041             \\
16  &          &          1,221,281            &      293,665               &      270,481                \\
32  &          &           4,877,633        &           1,171,009          &         1,078,561             \\
\bottomrule                     
\end{tabular}
}
\end{table}

A total of 800 samples were generated, of which 650 were used for training and 150 for testing (validation).
A data-driven loss function was used to compare the network output with the actual results from the numerical simulation training data.
The mean squared error is defined as

\begin{equation}
\label{eqn:cnn_loss}
MSE =  \frac{1}{N_{data} } \sum_{i}^{N_{data}} \left( T^i_{Pred} - T^i_{Target} \right)^{2},
\end{equation}

where $T^i_{Pred}$ is the temperature prediction for sample $i$, $T^i_{Target}$ is the known solution, and $N_{data}$ is the total number of data samples used for either training or testing.

The neural networks were all built using PyTorch 
and trained using an NVIDIA GeForce RTX 3090 GPU with 24Gb RAM at the University of Stuttgart.
The training and testing data, as well as the Python code for the network is provided in the DaRUS dataset "Replication Data for: Geothermal-ML - predicting thermal plume from groundwater heat pumps" \cite{darus-31842022}.

\section{Results and Discussion}

The following section provides the results and discussion for all three networks: TurbNet-Geo (TNG), TurbNet-Geo-Light (TNG-L) and TurbNet-Geo-NoSkip-Light (TNG-NS-L).
The training and testing loss for the three network architectures and varying initial features are shown in table \ref{tab:extended_loss_results} at the 50,000$^{th}$ epoch.  

\begin{table}[!ht]
  \centering
  \caption{\small{Training and testing loss of all networks with varying number of initial features. The three networks were trained with 4, 8, 16 and 32 initial features ('Init. Feat.'), which varies the total number of training parameters in the network. The training loss ('Loss') and the testing loss ('Test Loss') are evaluated at the 50,000$^{th}$ epoch. }}
  \label{tab:extended_loss_results}
  \small{
\begin{tabular}{lllllllll}
\toprule
   & \multicolumn{2}{c}{TurbNetGeo}             & \multicolumn{1}{c}{} & \multicolumn{2}{c}{TurbNetGeo-Light}      & \multicolumn{1}{c}{} & \multicolumn{2}{c}{TurbNetGeo-NoSkip-Light}                                    \\
Init. Feat.   & \multicolumn{1}{c}{Loss}                 & \multicolumn{1}{c}{Test Loss} & \multicolumn{1}{c}{} & \multicolumn{1}{c}{Loss}                 & \multicolumn{1}{c}{Test Loss} & \multicolumn{1}{c}{} & \multicolumn{1}{c}{Loss}                & \multicolumn{1}{c}{Test Loss} \\
   \midrule
4 & 1.19 $\cdot 10^{-5}$  & 0.0659  &   & 1.38 $\cdot 10^{-5}$ & 0.0556    &     & 2.40 $\cdot 10^{-5}$ & 0.0224             \\
8 & 6.22 $\cdot 10^{-6}$ & 0.0455   &   & 3.93 $\cdot 10^{-6}$   & 0.0423    &     & 1.26 $\cdot 10^{-6}$ & 0.0173  \\
16 & 2.22 $\cdot 10^{-7}$  & 0.0812  &   & 3.31 $\cdot 10^{-7}$ & 0.0318    &     & 3.96 $\cdot 10^{-7}$ & 0.0171             \\
32 & 2.16 $\cdot 10^{-7}$ & 0.0433   &   & 2.17 $\cdot 10^{-7}$   & 0.0270    &     & 2.66 $\cdot 10^{-7}$ & 0.0159  \\
\bottomrule                     
\end{tabular}
}
\end{table}

The training loss for all three networks with 4 and 8 initial features are at least an order of magnitude larger than with 16 and 32 initial features. 
The TNG-NS-L has the largest training loss at 50,000 epochs for both 16 and 32 initial features. 
The TNG network has the lowest training loss for both 16 and 32 initial features, but only marginally lower with 4 times the number of trainable parameters than the other two networks, making each network evaluation more expensive. 
Despite having fewer trainable parameters than the other networks, the TNG-NS-L has the lowest testing loss.
Overall, the three network architectures show similar training losses for the same number of initial features.
This indicates that even the smallest network architecture is already expressive enough to capture the prediction task at hand.

\begin{figure*}[t!]
\centering
    \includegraphics[width=0.8\textwidth]{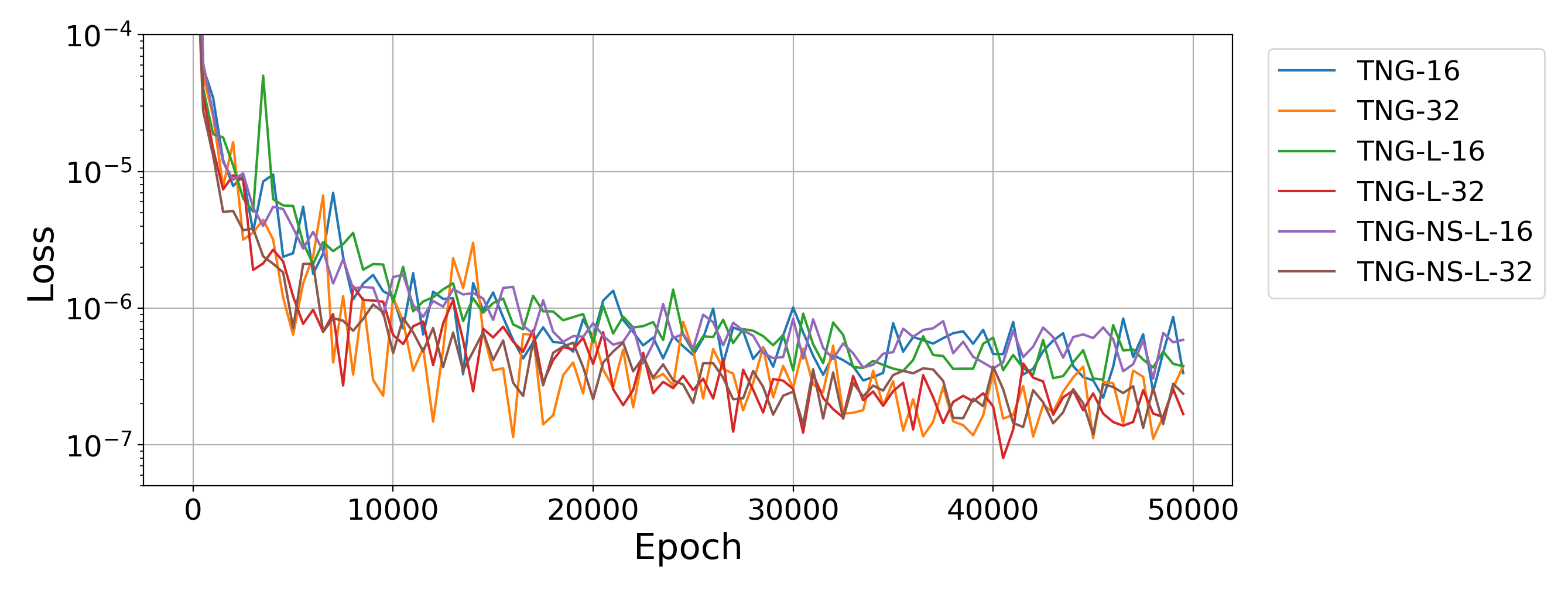}
    \caption{\label{fig:training_loss} \small{Training loss for all three network architectures TurbNet-Geo (’TNG’), TurbNet-Geo-Light (’TNG-Light’) and TurbNet-NoSkip-Light (’TNG-NoSkip’), were evaluated for 16 and 32 initial features, denoted as ’-16’ and ’-32’, respectively. The training loss is plotted at every 200$^{th}$ epoch.}
    }
\end{figure*}

The training loss across all 50,000 epochs for all three network architectures with 16 and 32 initial features is shown in Figure \ref{fig:training_loss}.   
For each case, the loss value is only plotted for every 500$^{th}$ epoch, with no smoothing performed on the data.  
Examining the training loss over time in Figure \ref{fig:training_loss}, there is little difference between the networks with an equivalent number of 16 or 32 initial features. 
It is clear that the number of initial features influences the training loss more than the number of network layers. 

\begin{figure*}
\centering
    \includegraphics[width=0.8\textwidth]{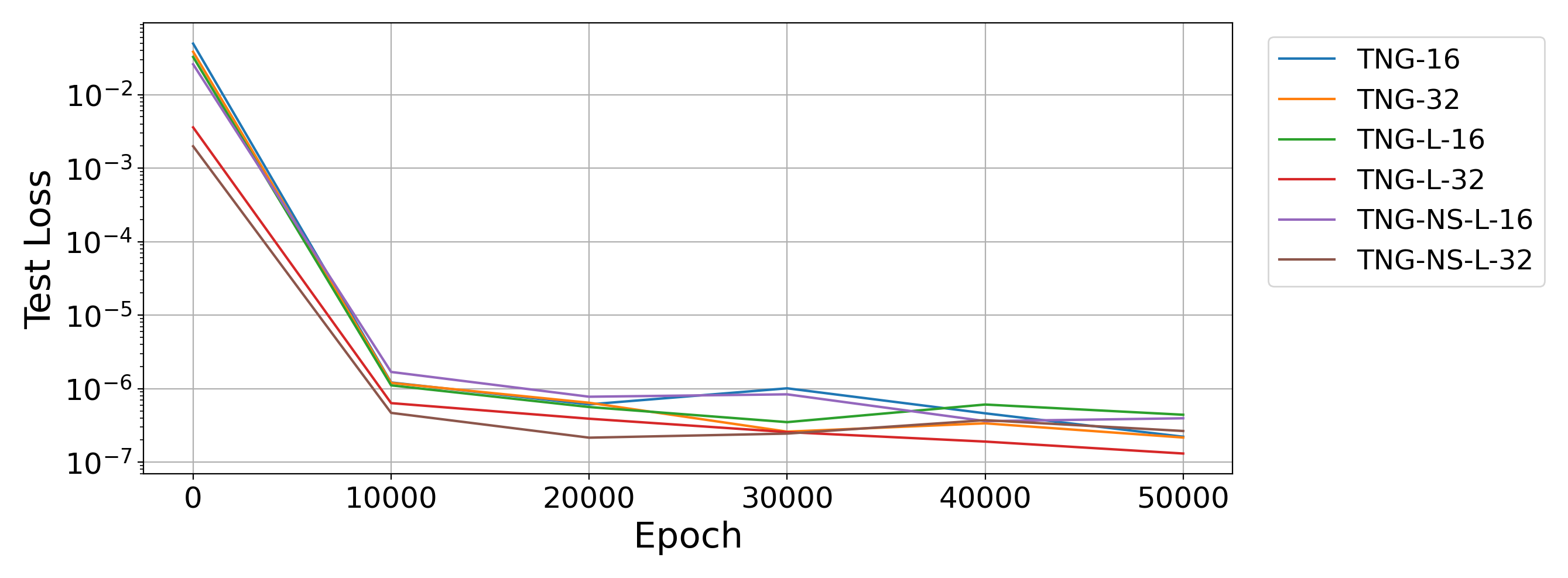}
    \caption{\label{fig:testing_loss}\small{Testing loss for all three network architectures TurbNet-Geo (’TNG’), TurbNet-Geo-Light (’TNG-Light’) and TurbNet-NoSkip-Light (’TNG-NoSkip’), were evaluated for 16 and 32 initial features, denoted as ’-16’ and ’-32’, respectively. The testing loss is plotted at every 5000$^{th}$ epoch.}}
\end{figure*}

The testing loss, measured every 10,000$^{th}$ epoch, is shown for all three network architectures with 16 and 32 initial features in Figure \ref{fig:testing_loss}.   
Minor, if any, improvement in the testing loss is observed after the first test at 10,000 epochs, indicating that additional training does not aid in improving the real-world capability of the network on predicting the thermal plume on unseen Darcy velocity data. 
This would perhaps require more training and testing data to improve, or a fundamental change to the network is required to reduce the testing loss. 

For each network, a total of 150 test samples were evaluated and categorized according to the maximum error $\epsilon_{max}$ across all pixels:
\begin{enumerate}[label=(\roman*)]
    \item good: $\epsilon_{max} <$ 0.5
    \item medium: 0.5 $< \epsilon_{max} <$ 1.0
    \item bad: 1.0 $< \epsilon_{max}$
\end{enumerate}
The total number of predictions categorized as "good", "medium" and "bad" for each network and varying number of initial features, is shown in table \ref{tab:number_gmb}.
The TNG network performs only moderately better than TNG-L, with 24 and 22 "good" prediction samples, respectively.
The TNG-NS-L has 15 "good" predictions, while the TNG is a might larger network with more trainable parameters. 
To reduce the network size as much as possible while maintaining reasonably good predictions, we focus on the TNG-L and TNG-NS-L networks as the preferred networks for further analysis. 
Comparing these two, the TNG-L has more test samples categorized as good for 16 and 32 initial features and more samples categorized as medium for 16 initial features. 
Even though the testing loss is lower for the TNG-NS-L network, there may be few pixels where the error is large, forcing test samples to be categorized as "medium" or "bad". 
However, we select the TNG-L with 32 initial features as the current default network for the rest of the analysis and compare the analysis to this network.

\begin{table}[b!]
\centering
\caption{\small{Number of samples classified as 'good', 'medium' and 'bad' predictions for all three network architectures. Any prediction with the maximum absolute error $\vert \epsilon_{max} \vert $< 0.5$\degree$C is defined as 'good', any prediction with 0.5$\degree$C < $\vert \epsilon_{max} \vert $ < 1$\degree$C is defined as 'medium' and all others are defined as bad.}}
  \label{tab:number_gmb}
  \small{
\begin{tabular}{l|ccc|ccc|ccc}
\toprule
            & \multicolumn{3}{c}{\textbf{TurbNetGeo}} & \multicolumn{3}{c}{\textbf{TurbNetGeo-Light}} & \multicolumn{3}{c}{\textbf{TurbNetGeo-NoSkip-Light}} \\ \cline{2-10} 
Init. Feat. & Good        & Medium         & Bad         & Good        & Medium         & Bad         & Good        & Medium         & Bad           \\
\midrule
4           & 0        & 5         & 145       & 0          & 3         & 147        & 0           & 1           & 149         \\
8                    & 1        & 50        & 99        & 1          & 28        & 121        & 1           & 43          & 106         \\
16                   & 7        & 81        & 62        & 14         & 56        & 80         & 8           & 64          & 78          \\
32                   & 24       & 66        & 60        & 22         & 63        & 65         & 15          & 69          & 66         \\
\bottomrule    
\end{tabular}
}
\end{table}

\begin{figure*}[t!]
    \centering
    \begin{subfigure}[b]{0.24\linewidth}
    \centering
    \includegraphics[width=\textwidth]{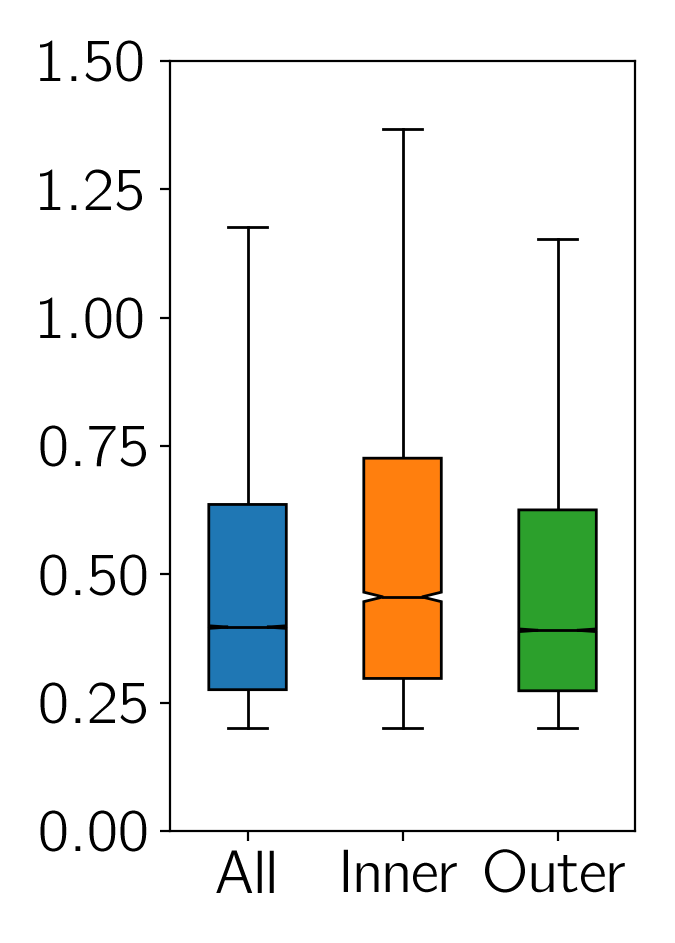}
     \caption{\small{TNG-L-16}}
     \end{subfigure}%
     \hspace{0.05cm}
     \begin{subfigure}[b]{0.24\linewidth}
     \centering
     \includegraphics[width=\textwidth]{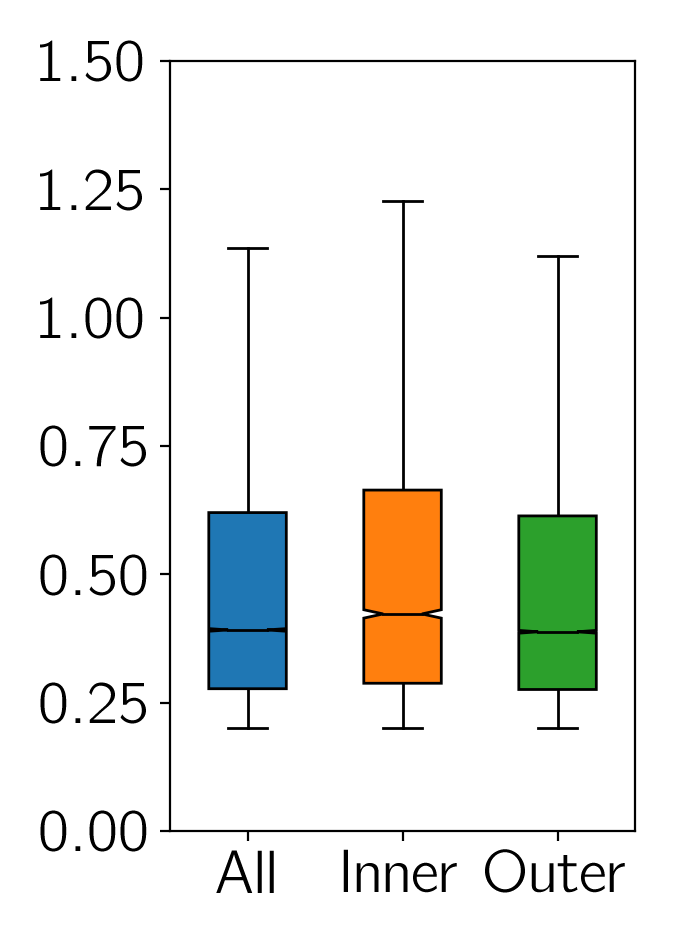}
     \caption{\small{TNG-L-32}}
     \end{subfigure}%
     \hspace{0.05cm}
      \begin{subfigure}[b]{0.24\linewidth}
    \centering
    \includegraphics[width=\textwidth]{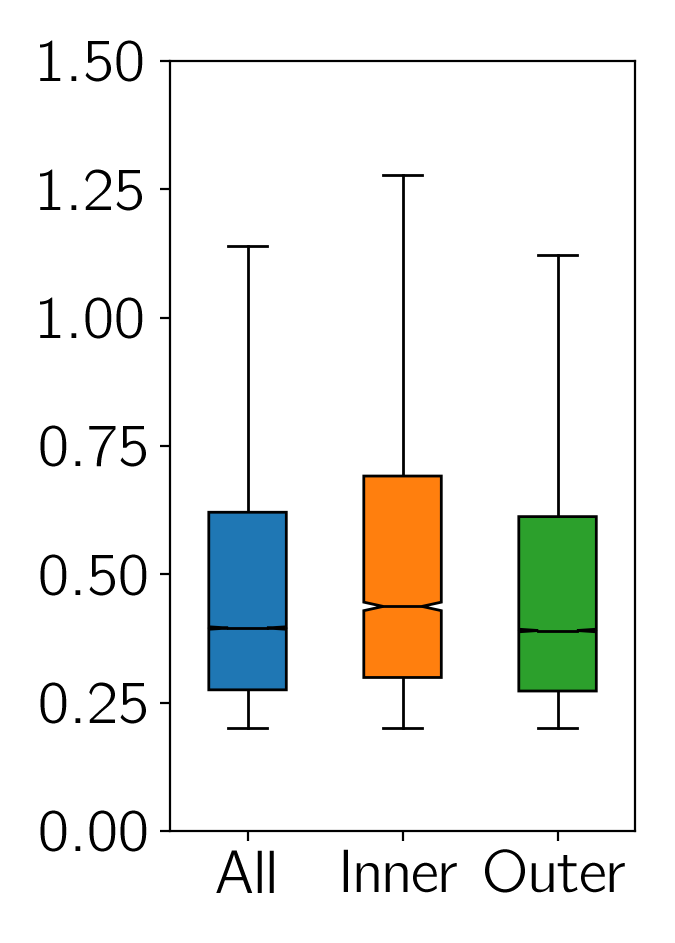}
     \caption{\small{TNG-NS-L-16}}
     \end{subfigure}%
     \hspace{0.05cm}
     \begin{subfigure}[b]{0.24\linewidth}
     \centering
     \includegraphics[width=\textwidth]{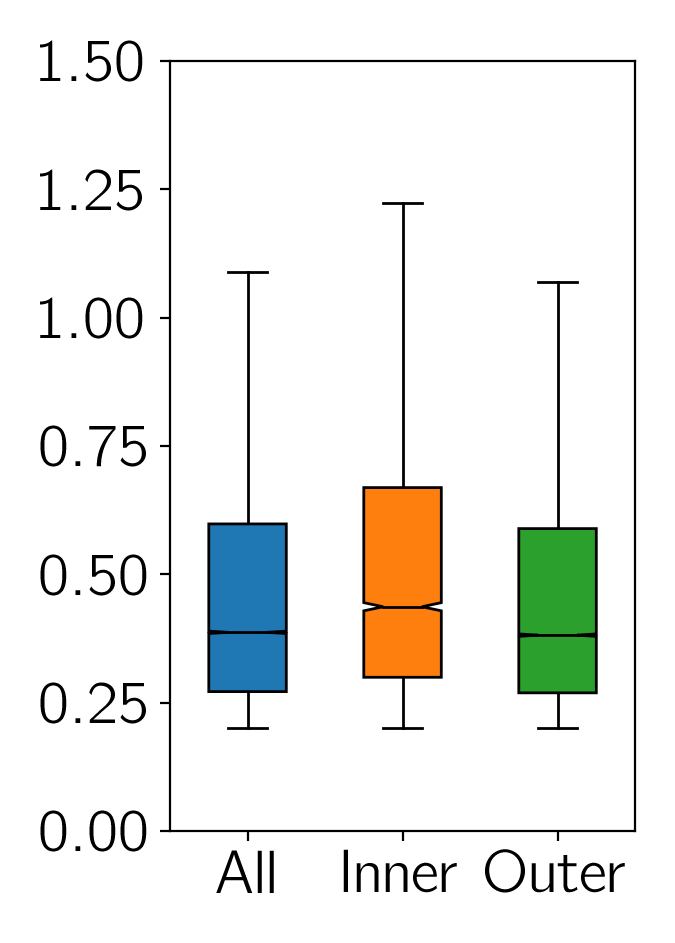}
     \caption{\small{TNG-NS-L-16}}
     \end{subfigure}%
     \caption{\small{Per-pixel error magnitude box plots for the TurbNet-Geo-Light and TurbNet-Geo-NoSkip-Light network architectures with 16 and 32 initial features for all 150 test samples. Each box provides the median $Q_2$, lower quartile $Q_1$ and upper quartile $Q_3$. The top and bottom edges (whiskers) define the 1.5 interquartile range (1.5$\cdot (Q_3$ - $Q_1 )$) values of the error magnitude. Only error values larger than 0.2$\degree$C were added to the dataset for the box plots.
     }}
     \label{fig:boxplot}
  \end{figure*}

Table \ref{tab:number_gmb} categorizes the predictions for the test set into "good", "medium" and "bad" predictions based on the maximum point-wise error.
For the TurbNetGeo-Light (TNG-L) and TurbNetGeo-NoSkip-Light (TNG-NS-L) networks with 16 and 32 initial features, box plots (box and whisker plot) of the per-pixel error magnitude across all 150 test samples are plotted in Figure \ref{fig:boxplot}(A) - (D).  
Each box provides the median $Q_2$ (middle black line), lower quartile $Q_1$ and upper quartile $Q_3$ (box edges). The top and bottom edges (whiskers) define the 1.5$\times$ interquartile range (IQR), i.e., 1.5$\cdot (Q_3$ - $Q_1 )$ values of the error magnitude.  
For each test (A) to (D), the box plots were generated using values where the error was more than 0.2$\degree$C to ignore minor fluctuations of the prediction output error in regions where the background temperature was 10$\degree$C, i.e., to ignore errors far away from the thermal plume. 
Furthermore, all points within an inner region of 7 pixels from the heat pump injection site in each direction are shown in the 'Inner' plots and all other points are shown in the 'Outer' plots. 
This highlights if the error is, in general, larger near the heat pump injection site or further downstream the thermal plume. 
For all box plots, the median error is below 0.4$\degree$C, indicating that at least half of the pixels where the error is above 0.2$\degree$C is also below 0.5$\degree$C (the "good" criterion). 
The largest upper interquartile point occurs for the TNG-L-16 'Inner' with a magnitude of 0.73$\degree$C, meaning 75\% of error values are beneath this value. 
As these plots also include the "bad" predictions, this shows that the networks are able to provide reasonably good predictions, and only few pixels have a large error that cause the "bad" categorization.
The 'Inner' box plots have a larger interquartile range and top whisker than the 'All' plot as the error tends to be slightly larger in the middle of the domain, leaving fewer small error values to pull the box plot down. 

\begin{figure}[ht!]
    \centering
    \begin{subfigure}[b]{0.95\linewidth}
        \includegraphics[width=\textwidth]{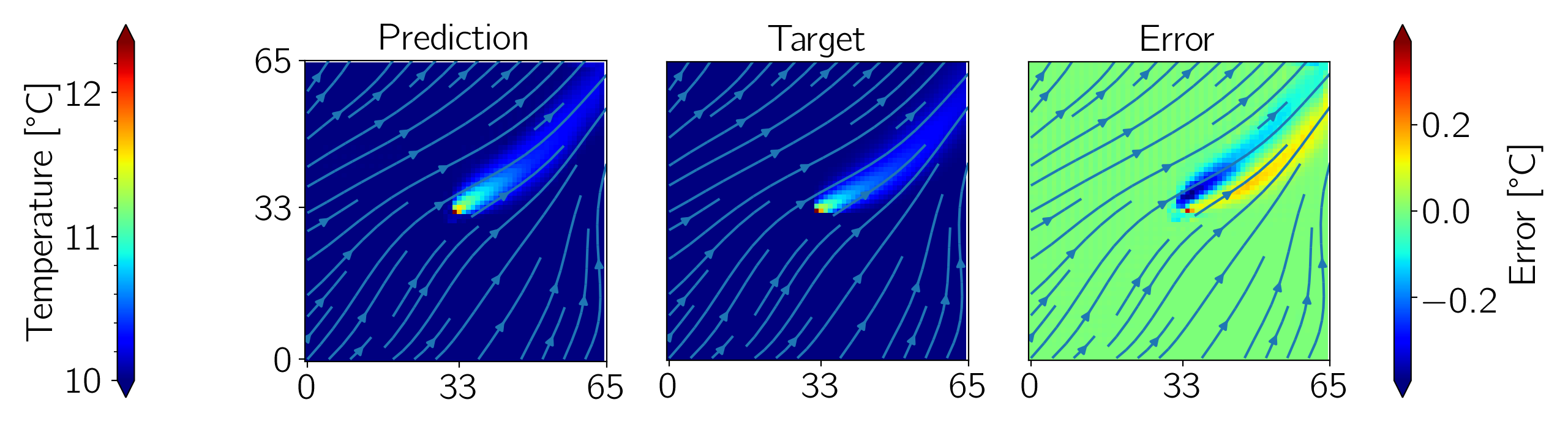}
     \end{subfigure}%
     \\
    \begin{subfigure}[b]{0.95\linewidth}
        \includegraphics[width=\textwidth]{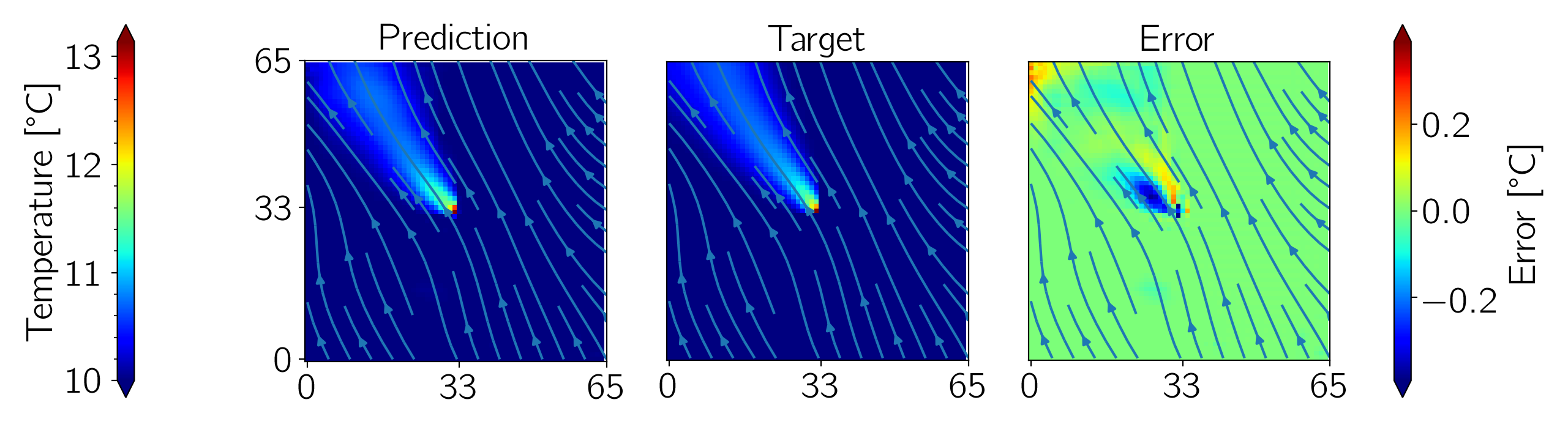}
     \end{subfigure}%
    \\
    \begin{subfigure}[b]{0.95\linewidth}
        \includegraphics[width=\textwidth]{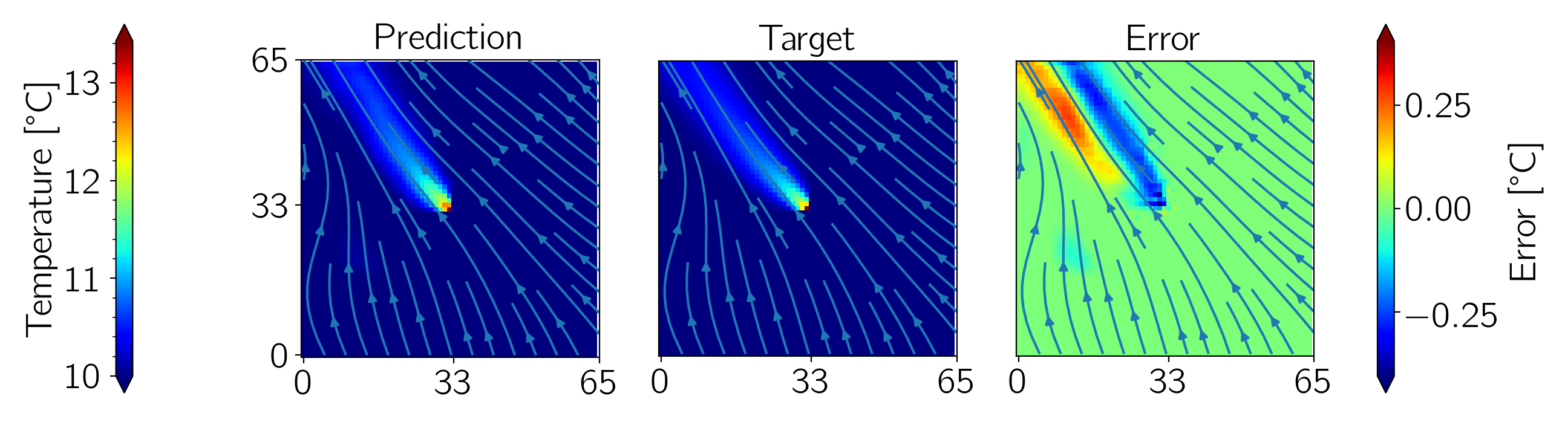}
     \end{subfigure}%
    \\
    \begin{subfigure}[b]{0.95\linewidth}
        \includegraphics[width=\textwidth]{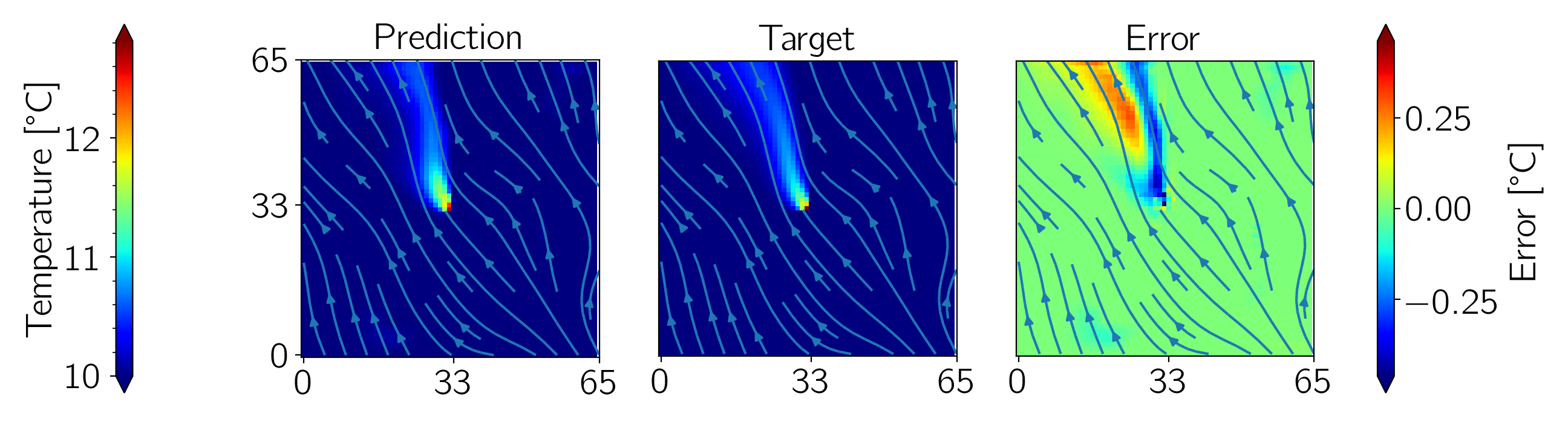}
     \end{subfigure}%
    \caption{\small{TurbNet-Geo-Light: "good" network prediction for four test samples with 32 initial features. The test sample numbers listed from top to bottom: 35, 96, 100 and 136.
     }}
    \label{fig:good_TNG_L_32}
  \end{figure}

\begin{figure}[ht!]
    \centering
    \begin{subfigure}[b]{0.95\linewidth}
        \includegraphics[width=\textwidth]{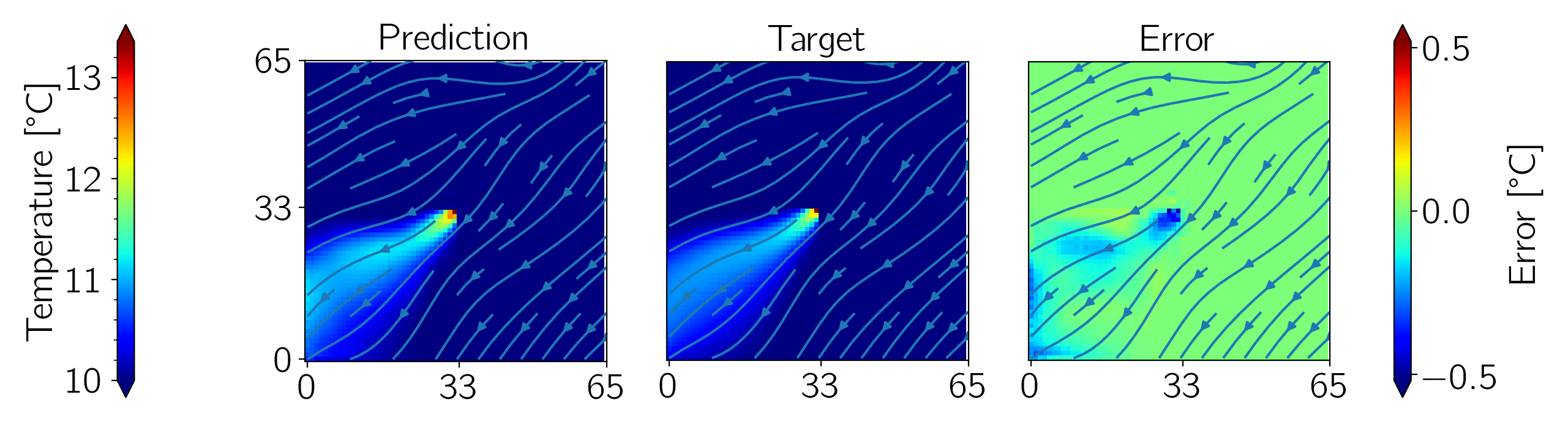}
     \end{subfigure}%
     \\
    \begin{subfigure}[b]{0.95\linewidth}
        \includegraphics[width=\textwidth]{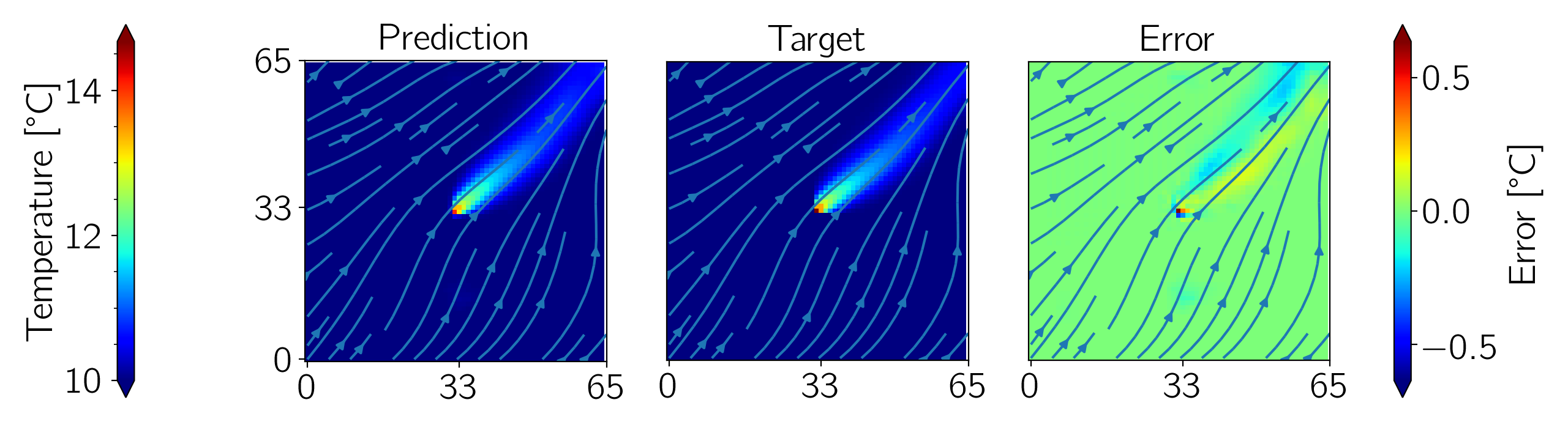}
     \end{subfigure}%
    \\
    \begin{subfigure}[b]{0.95\linewidth}
        \includegraphics[width=\textwidth]{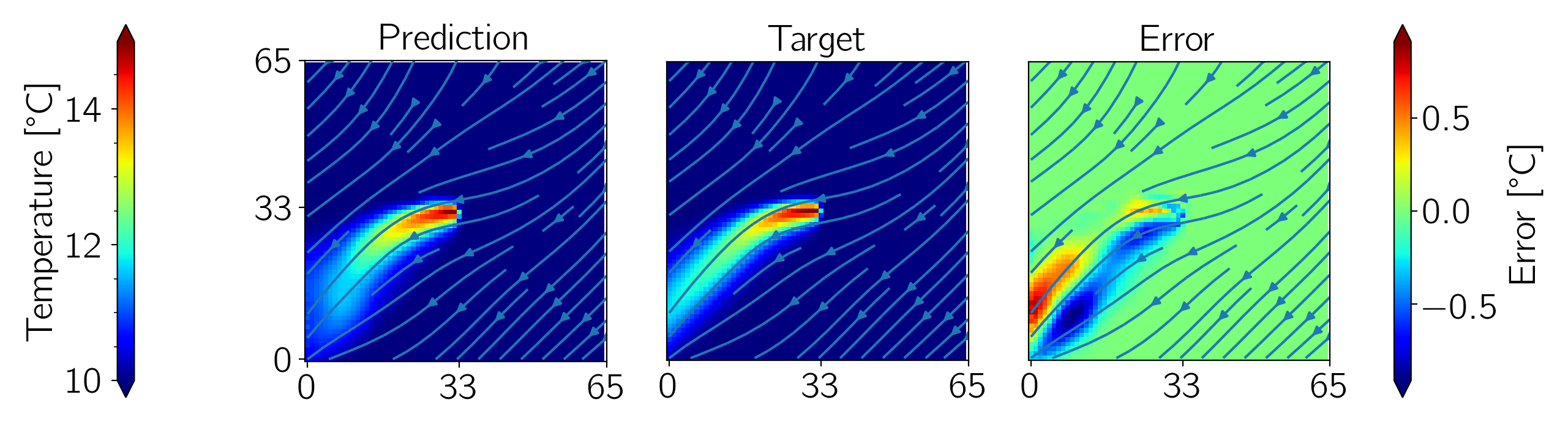}
     \end{subfigure}%
    \\
    \begin{subfigure}[b]{0.95\linewidth}
        \includegraphics[width=\textwidth]{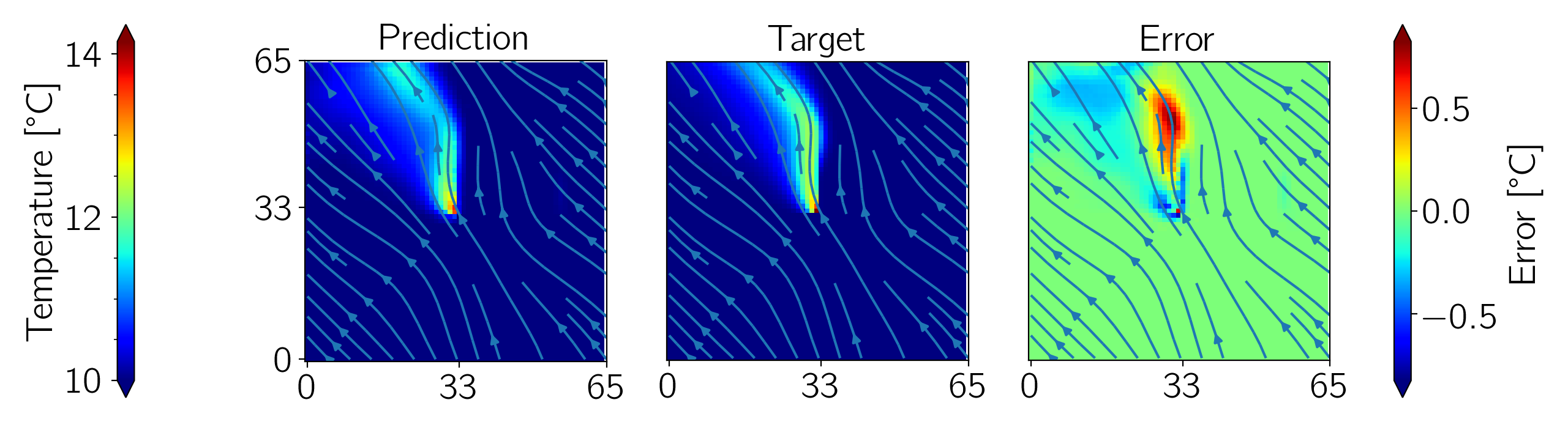}
     \end{subfigure}%
     \caption{\small{TurbNet-Geo-Light: "medium" network prediction for four test samples with 32 initial features.  The test sample numbers listed from top to bottom: 16, 76, 99 and 12. 
     }}
    \label{fig:medium_TNG_L_32}
  \end{figure}

\begin{figure}[ht!]
    \centering
    \begin{subfigure}[b]{0.95\linewidth}
        \includegraphics[width=\textwidth]{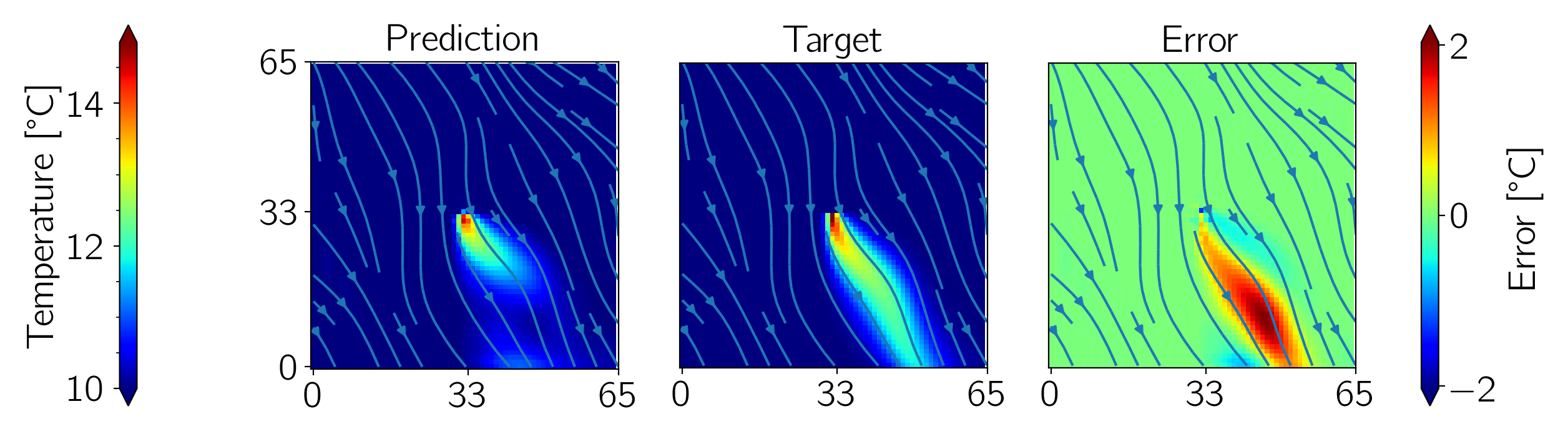}
     \end{subfigure}%
     \\
    \begin{subfigure}[b]{0.95\linewidth}
        \includegraphics[width=\textwidth]{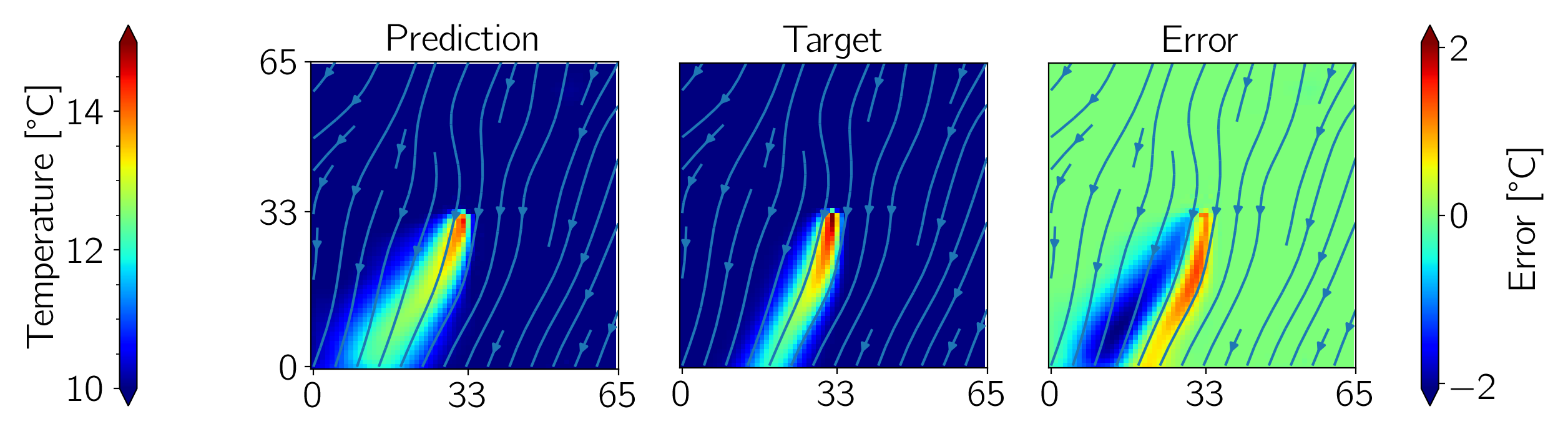}
     \end{subfigure}%
    \\
    \begin{subfigure}[b]{0.95\linewidth}
        \includegraphics[width=\textwidth]{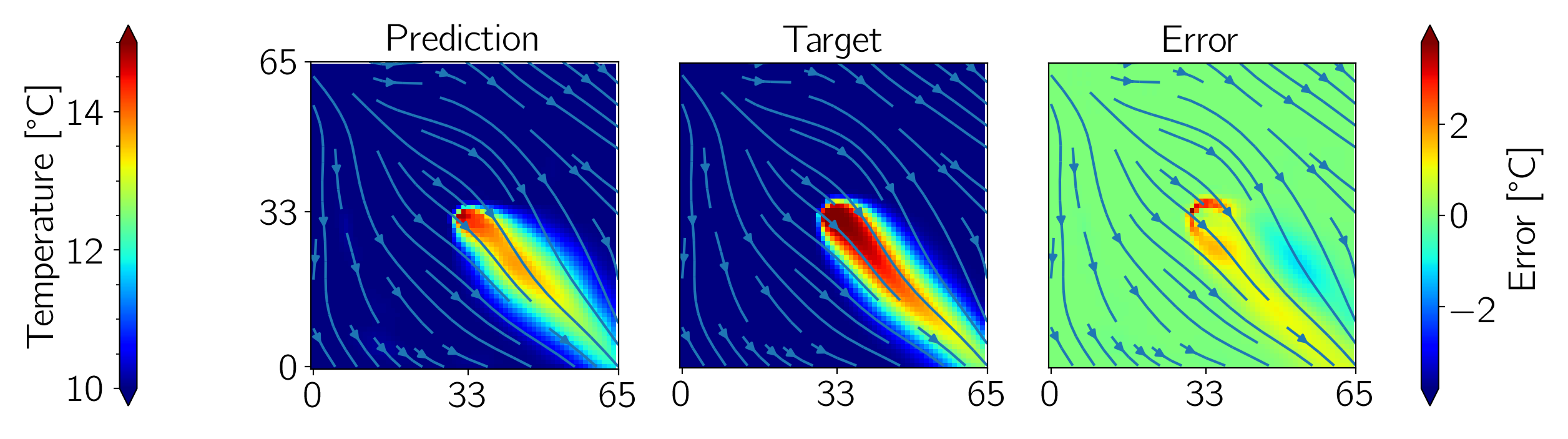}
     \end{subfigure}%
    \\
    \begin{subfigure}[b]{0.95\linewidth}
        \includegraphics[width=\textwidth]{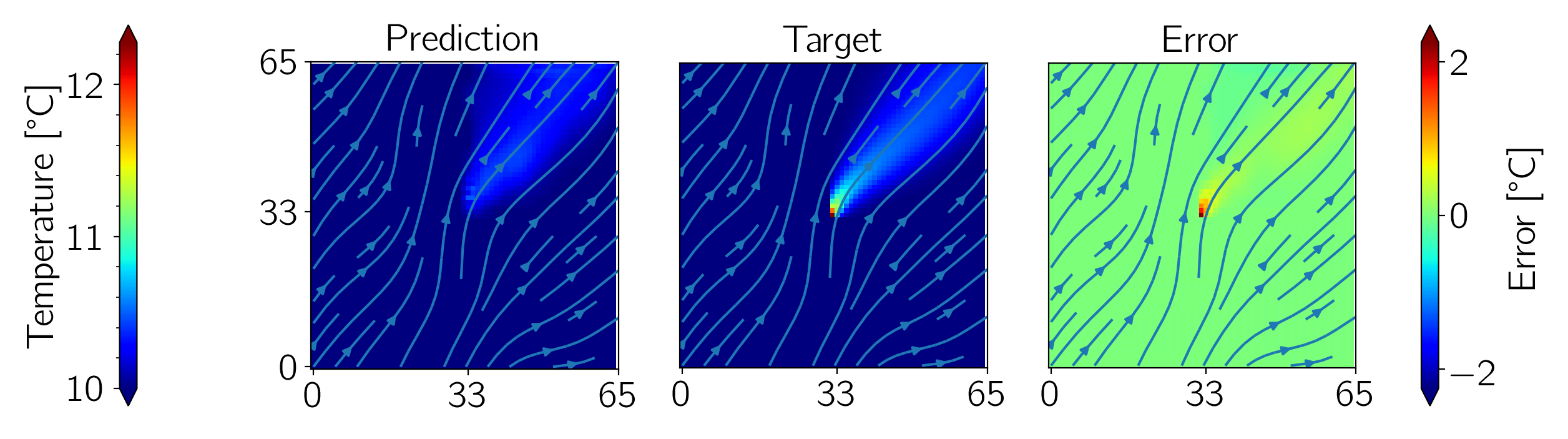}
     \end{subfigure}%
    \caption{\small{TurbNet-Geo-Light: "bad" network prediction for four test samples with 32 initial features.  The test sample numbers listed from top to bottom: 20, 52, 66 and 139. 
     }}
    \label{fig:bad_TNG_L_32}
  \end{figure}

\begin{figure}[ht!]
    \centering
    \begin{subfigure}[b]{0.95\linewidth}
        \includegraphics[width=\textwidth]{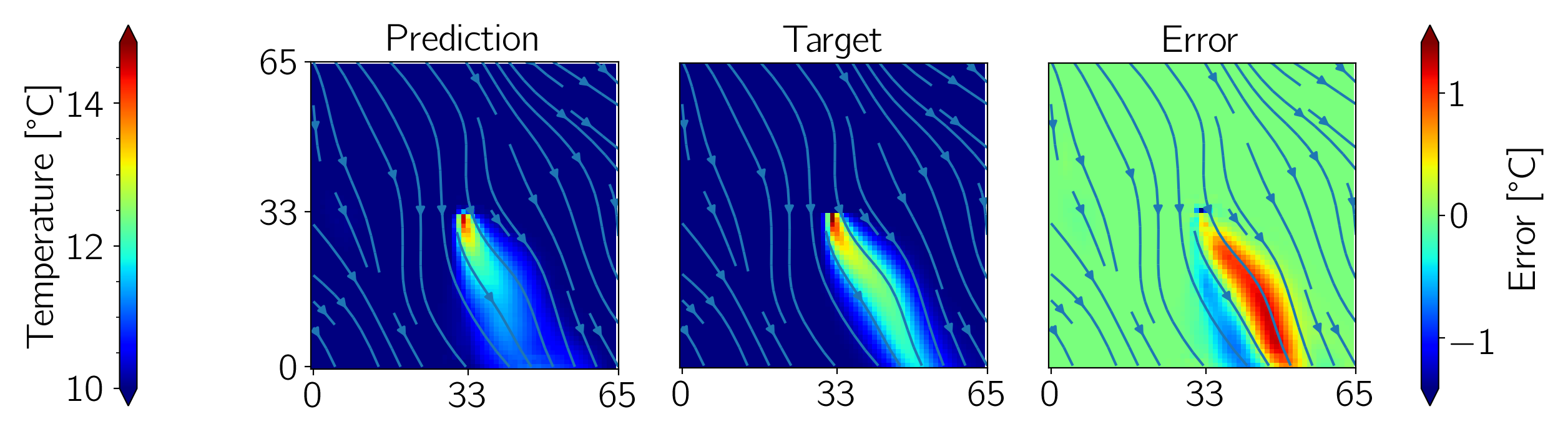}
        \caption{TNG-NS-L: sample 20}
     \end{subfigure}%
     \\
    \begin{subfigure}[b]{0.95\linewidth}
        \includegraphics[width=\textwidth]{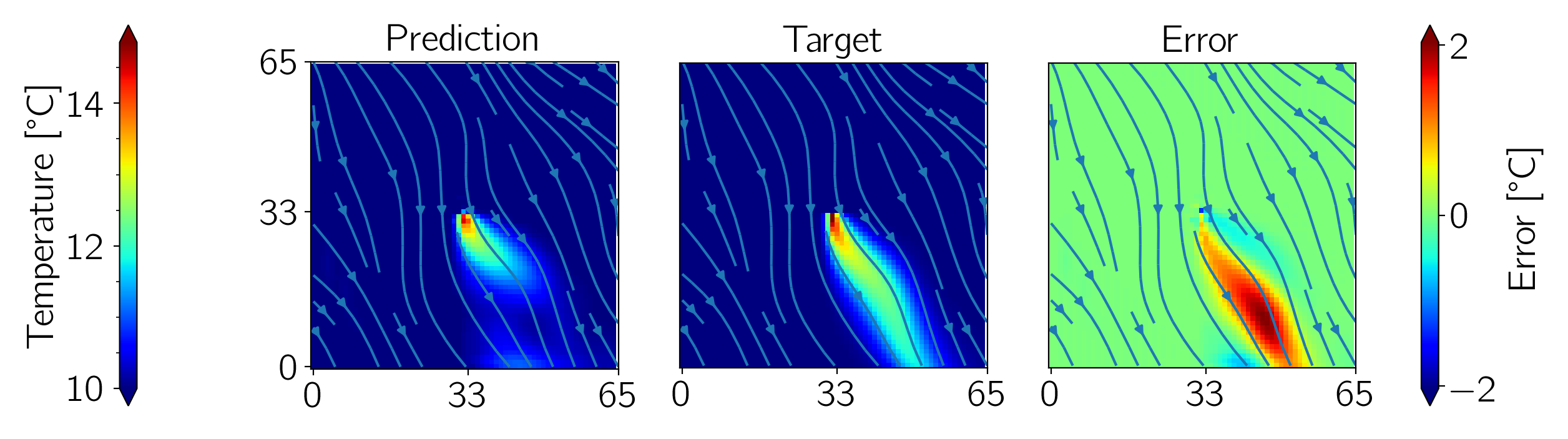}
        \caption{TNG-L: sample 20}
     \end{subfigure}%
    \\
    \begin{subfigure}[b]{0.95\linewidth}
        \includegraphics[width=\textwidth]{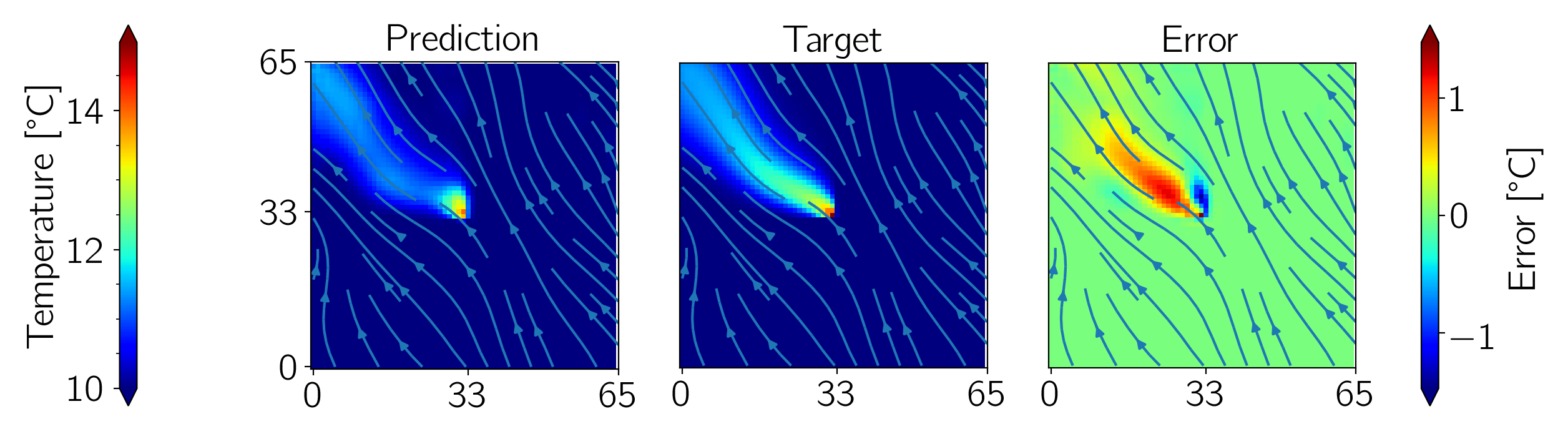}
        \caption{TNG-NS-L: sample 51}
     \end{subfigure}%
    \\
    \begin{subfigure}[b]{0.95\linewidth}
        \includegraphics[width=\textwidth]{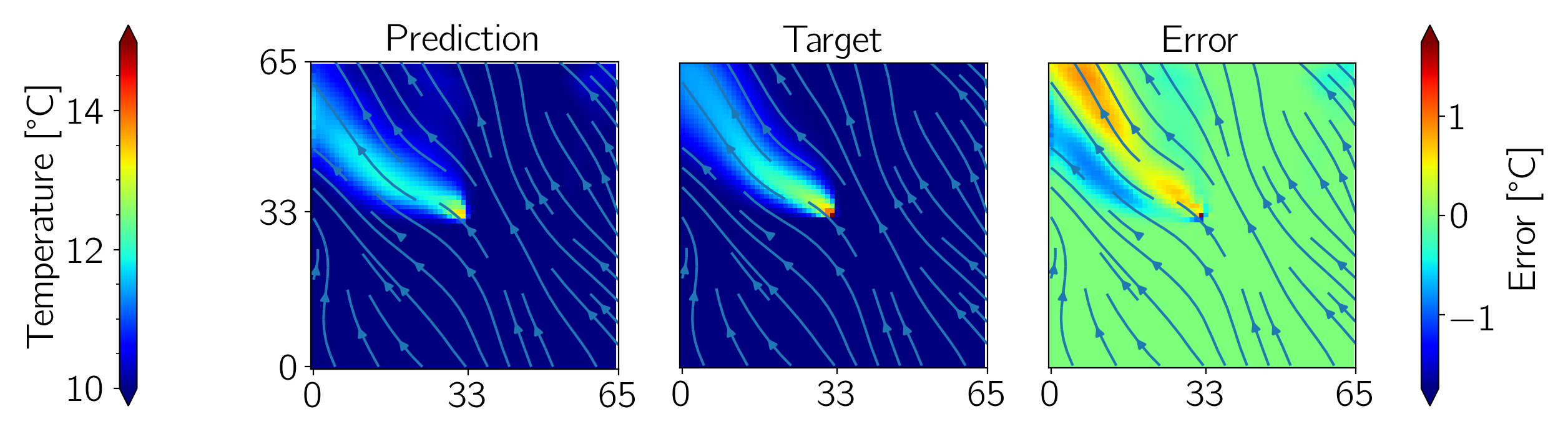}
        \caption{TNG-L: sample 51}
     \end{subfigure}%
     \caption{\small{TurbNet-Geo-Light versus TurbNet-Geo-NoSkip-Light: "bad" network prediction for two test samples with 32 initial features. 
     }}
    \label{fig:comp_1}
  \end{figure}

\begin{figure}[ht!]
    \centering
    \begin{subfigure}[b]{0.95\linewidth}
        \includegraphics[width=\textwidth]{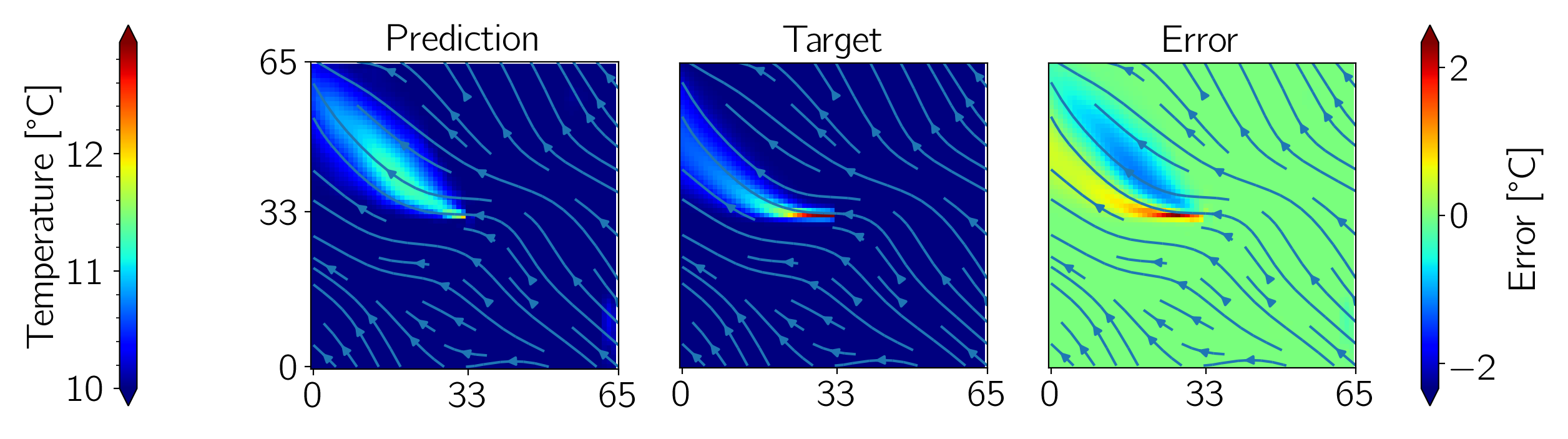}
        \caption{TNG-NS-L: sample 137}
     \end{subfigure}%
     \\
    \begin{subfigure}[b]{0.95\linewidth}
        \includegraphics[width=\textwidth]{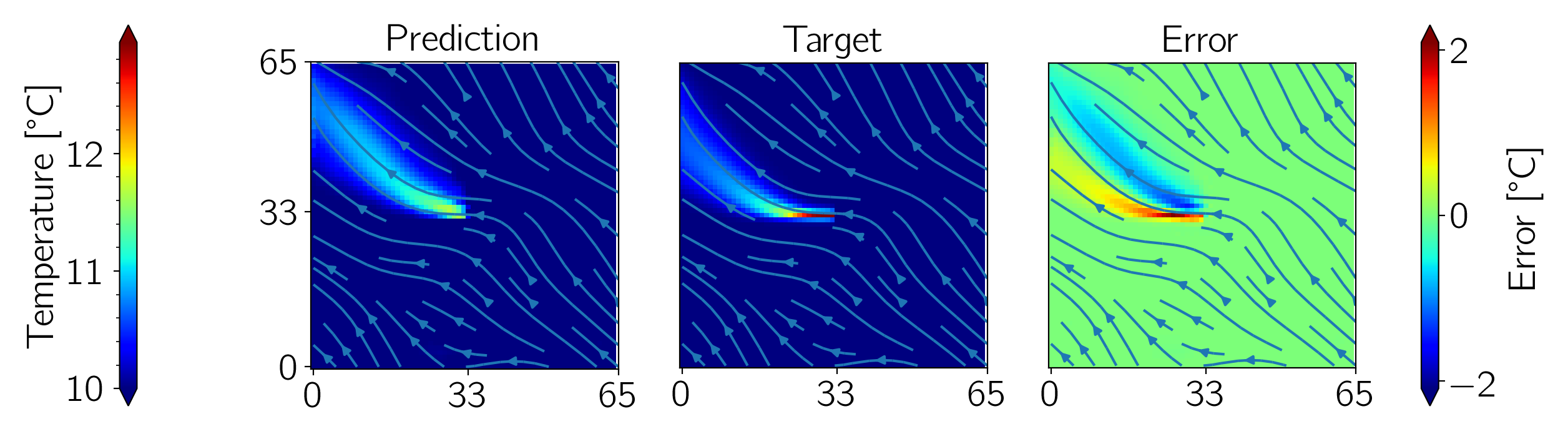}
        \caption{TNG-L: sample 137}
     \end{subfigure}%
    \\
    \begin{subfigure}[b]{0.95\linewidth}
        \includegraphics[width=\textwidth]{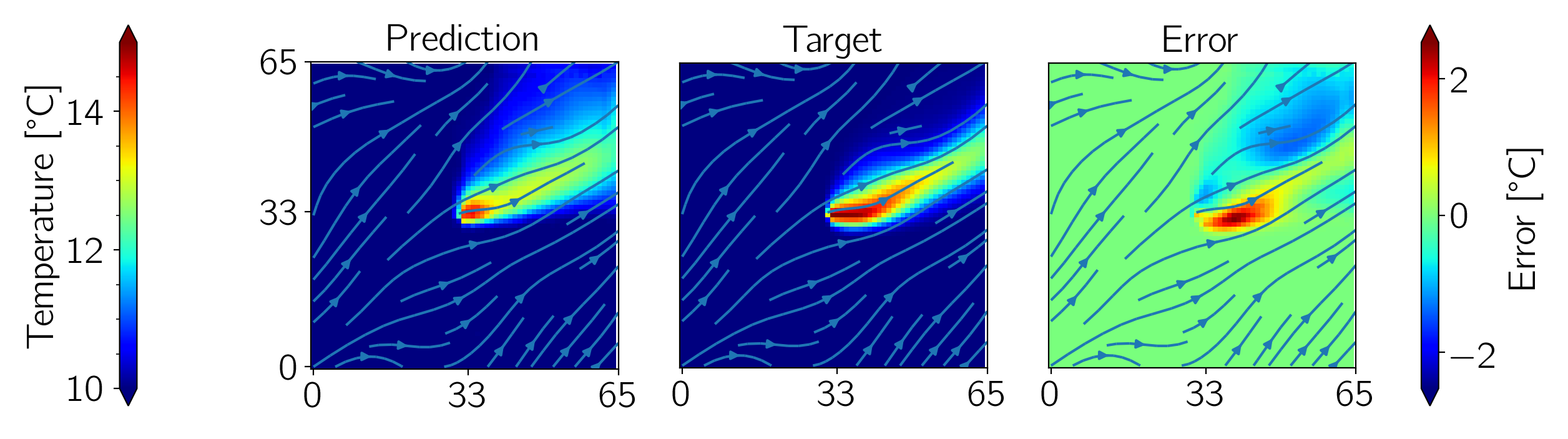}
        \caption{TNG-NS-L: sample 50}
     \end{subfigure}%
    \\
    \begin{subfigure}[b]{0.95\linewidth}
        \includegraphics[width=\textwidth]{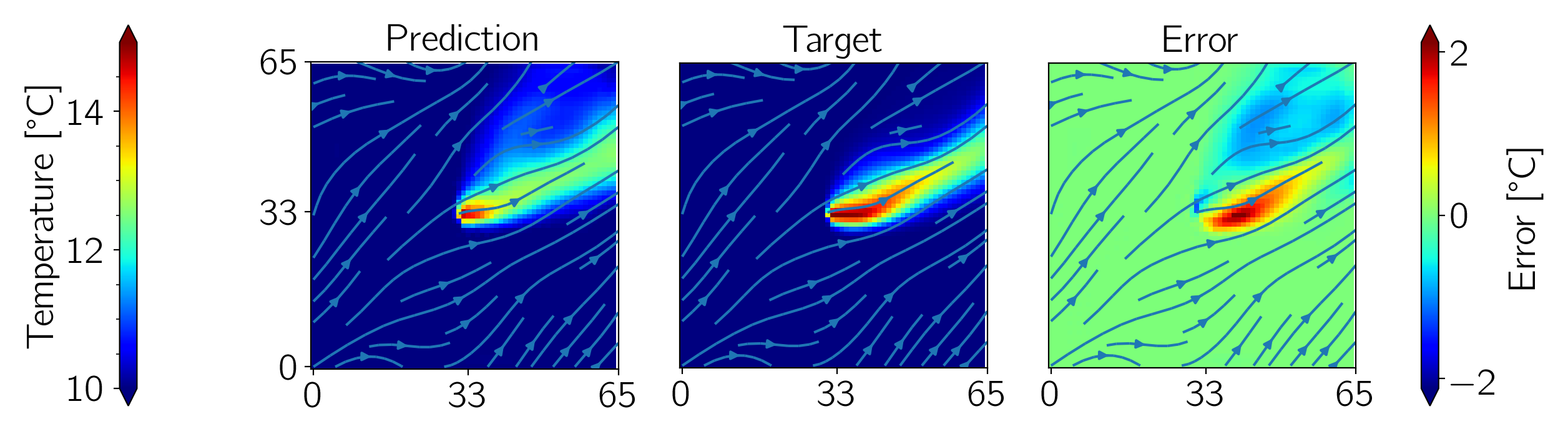}
        \caption{TNG-L: sample 50}
     \end{subfigure}%
    \caption{\small{TurbNet-Geo-Light versus TurbNet-Geo-NoSkip-Light: "bad" network prediction for two samples with 32 initial features. 
     }}
    \label{fig:comp_2}
  \end{figure}

Examining the training and testing loss conveys how well the network performs overall. 
Additionally, the box plots tells use how many pixels are considered good, medium and bad. 
However, it does not indicate where or how the greatest errors occur, i.e., whether it occurs 
at the GWHP location itself or if the thermal plume is unable to follow the streamlines. 
Therefore, a qualitative assessment of the network's performance is required to gain further insight. 
Four different network predictions for the TNG-L network with 32 initial features from the "good", 
"medium" and "bad" categories are shown in Figure \ref{fig:good_TNG_L_32}, Figure \ref{fig:medium_TNG_L_32} and Figure \ref{fig:bad_TNG_L_32}, respectively.
Finally, the difference between the TurbNetGeo-Light (TNG-L) and the TurbNetGeo-NoSkip-Light (TNG-NS-L) with 32 initial features from the "bad" category are exemplified on four test samples in Figure \ref{fig:comp_1} and Figure \ref{fig:comp_2}.
 

In Figure \ref{fig:good_TNG_L_32} the maximum error occurs at the heat pump location for all four samples.
However, the second sample (second row) has very small errors in the thermal plume in comparison to the others. 
The top and bottom samples indicate a good ability to follow the streamline as its path changes. 
The similar looking middle samples show how the thermal plume is able to spread out when the 
streamlines diverge (second row), whereas the thermal plume remains narrow when the streamlines remain straight (third row). 
As expected, the "good" category performs well in general. 

Four test samples from the "medium" category for the TNG-L with 32 initial features are shown in Figure \ref{fig:medium_TNG_L_32}. 
The top two samples have a very low error magnitude in the thermal plume, and was 
categorized into "medium" due to a small error spike at the heat pump location. 
The large error occurs when the GWHP location in the plume prediction is shifted one or two 
pixels compared to the target solution, shifting the maximum temperature pixel. 
If shifted in the wrong direction, i.e., if the plume extends downwards but the GWHP pixel 
location is shifted upwards, a large difference exists at this shifted location only and artificially inflates the maximum error value. 
However, the error values within the thermal plumes themselves are well within acceptable limits. 
The bottom two samples show that the plume can morph with the streamlines, but not well enough to keep the error below 0.5$\degree$C.

Four test samples from the "bad" category for the TNG-L with 32 initial features are shown in Figure \ref{fig:bad_TNG_L_32}. 
Each sample indicates various problems with some predictions. 
The top sample cannot accurately capture the temperature far downstream, whereas the second 
sample does not follow the streamline at all, but cuts across it instead. 
This is in contrast to our assumption that the thermal plume follows the streamline. 
The third sample cannot capture the high temperature at the heat pump location, where the 
plume is also relatively wide due to having a low Darcy velocity at the heat pump. 
This causes the higher temperature to diffuse outwards instead of being dragged downstream.  
Finally, the bottom sample is unable to capture the thermal profile near the heat pump. 

Another four test samples from the "bad" category for the TNG-NS-L (first and third image) and 
TNG-L (second and fourth image) with 32 initial features, comparing the ability of the two networks, 
are shown in Figure \ref{fig:comp_1} and Figure \ref{fig:comp_2}, each with two samples.  
For sample 20 in Figure \ref{fig:comp_1}, the TNG-NS-L network provides a better prediction than the TNG-L network, which cannot predict the entire plume correctly. 
The largest error occurs further downstream of the GWHP. 
However, the TNG-L network provides a better prediction of the downstream plume compared to the TNG-NS-L for sample 51. 

In Figure \ref{fig:comp_2}, the prediction for the first sample is similar for both networks, where the largest error is close to the GWHP. 
However, the TNG-NS-L network has a wider plume and does not quite follow the streamline as well as the TNG-L network. 
The final prediction is difficult for both networks. 
Both fail to predict the maximum temperature directly downstream of the GWHP and also over-predict the width of the plume. 

By comparing the two networks, both have their own positives and negatives. The TNG-NS-L sometimes 
outperforms the TNG-L, but it is not clear that it is necessarily better for all cases. 
The only clear benefit of the TNG-NS-L would be the simpler design and fewer trainable parameters for training. 
The provided examples are only a small set of samples that are available for comparison and were chosen 
from the set of 150 samples to highlight problems that may occur for the network prediction. 
Even for the "bad" category, the predictions are not completely unusable, but there are edge cases where 
the prediction would not be suitable to use in place of the high-fidelity solver. 
However, the CNN is suitable as an initial pre-processing step for the high-fidelity optimization problem, where highly interacting heat pumps can be identified. 
Therefore, the recommended network is the TNG-L with 32 initial features, followed by TNG-NS-L with 32 initial features. 
This network is suitable to implement inside the online evaluation tool of Fig  to complement the LAHM.


\vfill
\newpage

\vfill
\newpage

\section{Conclusion}

In this paper, we introduced a novel method for predicting the thermal plume downstream 
propagation from open-loop groundwater heat pump injection wells. 
Conventional methods for modeling heat pumps are either too expensive or lack a suitable level of accuracy. 
By understanding the governing equations for the subsurface water temperature and studying readily available simulation data, a seemingly simplistic convolutional neural network was built that accepts the 
subsurface Darcy velocities as input and outputs the thermal profile due to the presence of a heat pump. 
The network was based on a modified U-Net architecture to create three network variants, one comprised of 
6 layers with skip connections (TurbNet-Geo), one with 4 layers with skip connections (TurbNet-Geo-Light) and one with 4 layers without skip connections (TurbNet-NoSkip-Light). 

Training and testing data was generated for small 2D domains, by creating random permeability and pressure 
gradient boundary conditions, and performing numerical groundwater simulations with PFLOTRAN to obtain the Darcy velocities and temperature field.  
A total of 800 input-output samples were generated, with 650 samples used to train the network and 150 samples used to test the network prediction. 

The large TurbNet-Geo architecture was found to be only marginally more accurate than the smaller network architectures, 
but requiring nearly 4 times the number of trainable parameters. 
The prediction results were classified into good, medium and bad according to a specified criterion of $\vert \epsilon_{max} \vert <$ 0.5, 0.5 $< \vert \epsilon_{max} \vert <$ 1.0 and 1.0 $< \vert \epsilon_{max} \vert$, respectively. 
It was found that only a few pixels in the prediction caused the predictions to be classified as medium or bad. 
Notably, 75\% of all pixels where the error was larger than 0.2$\degree$C were also below 0.73$\degree$C, within the "medium" classification. 
Examining the thermal plume profiles of each network, each architecture showed good agreement between the prediction and the ground truth for most of the test samples.
There were a few samples where the network struggled to reasonably determine the shape of the plume following the streamlines.
However, this was limited to a few samples only and were presented in the discussion. 
Overall, the networks performed well as a fast surrogate model alternative to the high-fidelity solver. 
Future work will include training the network on larger datasets, larger domains, and more complex 
3D domains where generating the training data is significantly more expensive. 

\section{Funding}

Funded by Deutsche Forschungsgemeinschaft (DFG, German Research Foundation) under Germany’s Excellence Strategy—EXC 2075—390740016. 
We acknowledge the support by the Stuttgart Center for Simulation Science (SimTech).

\section{Conflicts of Interest}
The funders had no role in the design of the study, in the collection, analyses, or interpretation of data; in the writing of the manuscript, or in the decision to publish the results.

\section{References}
\bibliography{bibliography}

\begin{thebibliography}{10}

\bibitem{eu2030report}
Commission E, Centre JR, Economidou M, Ringel M, Valentova M, Zancanella P,
  et~al.
\newblock National energy and climate plans for 2021-2030 under the EU energy
  union : assessment of the energy efficiency dimension.
\newblock Publications Office; 2020.
\newblock Available from:
  \url{https://op.europa.eu/en/publication-detail/-/publication/2c751d22-4f07-11eb-b59f-01aa75ed71a1/language-en/format-PDF/source-271999318}.

\bibitem{Attard2020}
Attard G, Bayer P, Rossier Y, Blum P, Eisenlohr L.
\newblock {A novel concept for managing thermal interference between geothermal
  systems in cities}.
\newblock Renewable Energy. 2020;145:914-24.
\newblock Available from: \url{https://doi.org/10.1016/j.renene.2019.06.095}.

\bibitem{GarciaGil2020}
Garc{\'{i}}a-Gil A, Moreno MM, Schneider EG, Marazuela M{\'{A}}, Abesser C,
  L{\'{a}}zaro JM, et~al.
\newblock {Nested shallow geothermal systems}.
\newblock Sustainability (Switzerland). 2020;12(12).

\bibitem{Meng2019}
Meng B, Vienken T, Kolditz O, Shao H.
\newblock {Evaluating the thermal impacts and sustainability of intensive
  shallow geothermal utilization on a neighborhood scale : Lessons learned from
  a case study}.
\newblock Energy Conversion and Management. 2019;199(July):111913.
\newblock Available from: \url{https://doi.org/10.1016/j.enconman.2019.111913}.

\bibitem{Beck2013}
Beck M, Bayer P, de~Paly M, Hecht-M{\'{e}}ndez J, Zell A.
\newblock {Geometric arrangement and operation mode adjustment in low-enthalpy
  geothermal borehole fields for heating}.
\newblock Energy. 2013;49(1):434-43.
\newblock Available from: \url{http://dx.doi.org/10.1016/j.energy.2012.10.060}.

\bibitem{DePaly2012}
{De Paly} M, Hecht-M{\'{e}}ndez J, Beck M, Blum P, Zell A, Bayer P.
\newblock {Optimization of energy extraction for closed shallow geothermal
  systems using linear programming}.
\newblock Geothermics. 2012;43:57-65.
\newblock Available from:
  \url{http://dx.doi.org/10.1016/j.geothermics.2012.03.001}.

\bibitem{Cardoso2009}
Cardoso MA, Durlofsky LJ, Sarma P.
\newblock Development and application of reduced-order modeling procedures for
  subsurface flow simulation.
\newblock International Journal for Numerical Methods in Engineering.
  2009;77(9):1322-50.
\newblock Available from:
  \url{https://onlinelibrary.wiley.com/doi/abs/10.1002/nme.2453}.

\bibitem{Pasetto2011}
Pasetto D, Guadagnini A, Putti M.
\newblock POD-based Monte Carlo approach for the solution of regional scale
  groundwater flow driven by randomly distributed recharge.
\newblock Advances in Water Resources. 2011;34(11):1450-63.
\newblock Available from:
  \url{https://www.sciencedirect.com/science/article/pii/S0309170811001321}.

\bibitem{Li2013}
Li X, Chen X, Hu BX, Navon IM.
\newblock Model reduction of a coupled numerical model using proper orthogonal
  decomposition.
\newblock Journal of Hydrology. 2013;507:227-40.
\newblock Available from:
  \url{https://www.sciencedirect.com/science/article/pii/S0022169413006562}.

\bibitem{Vinuesa2022}
Vinuesa R, Brunton SL.
\newblock {Enhancing computational fluid dynamics with machine learning}.
\newblock Nat Comput Sci. 2022;2:358-66.
\newblock Available from: \url{https://doi.org/10.1038/s43588-022-00264-7}.

\bibitem{Chen2021}
Chen W, Wang Q, Hesthaven JS, Zhang C.
\newblock Physics-informed machine learning for reduced-order modeling of
  nonlinear problems.
\newblock Journal of Computational Physics. 2021;446:110666.
\newblock Available from:
  \url{https://www.sciencedirect.com/science/article/pii/S0021999121005611}.

\bibitem{Laubscher2017}
Laubscher R. Utilization of artificial neural networks to resolve chemical
  kinetics in turbulent fine structures of an advanced CFD combustion model;
  2017.

\bibitem{Toit2018}
{du Toit} P. An artificial intelligence approach for biomass devolatilisation
  in an industrial CFD model with advanced turbulence-chemistry interaction;
  2018.

\bibitem{Thuerey2019}
Thuerey N, Weißenow K, Prantl L, Hu X.
\newblock Deep learning methods for Reynolds-averaged Navier-Stokes simulations
  of airfoil flows.
\newblock American Institute of Aeronautics and Astronautics. 2019;58:686  707.

\bibitem{Wang2020}
Wang Y, Lin G.
\newblock Efficient deep learning techniques for multiphase flow simulation in
  heterogeneous porousc media.
\newblock Journal of Computational Physics. 2020;401:108968.
\newblock Available from:
  \url{https://www.sciencedirect.com/science/article/pii/S0021999119306734}.

\bibitem{Tang2021}
Tang M, Liu Y, Durlofsky LJ.
\newblock Deep-learning-based surrogate flow modeling and geological
  parameterization for data assimilation in 3D subsurface flow.
\newblock Computer Methods in Applied Mechanics and Engineering.
  2021;376:113636.
\newblock Available from:
  \url{https://www.sciencedirect.com/science/article/pii/S0045782520308215}.

\bibitem{Wang2021}
Gao H, Sun L, Wang JX.
\newblock {Super-resolution and denoising of fluid flow using physics-informed
  convolutional neural networks without high-resolution labels}.
\newblock Physics of Fluids. 2021;33(7).

\bibitem{Raissi2019}
Raissi M, Perdikaris P, Karniadakis GE.
\newblock Physics-informed neural networks: A deep learning framework for
  solving forward and inverse problems involving nonlinear partial differential
  equations.
\newblock Journal of Computational Physics. 2019;378:686-707.
\newblock Available from:
  \url{https://www.sciencedirect.com/science/article/pii/S0021999118307125}.

\bibitem{Zhu2019}
Zhu Y, Zabaras N, Koutsourelakis PS, Perdikaris P.
\newblock Physics-constrained deep learning for high-dimensional surrogate
  modeling and uncertainty quantification without labeled data.
\newblock Journal of Computational Physics. 2019;394:56-81.
\newblock Available from:
  \url{https://www.sciencedirect.com/science/article/pii/S0021999119303559}.

\bibitem{Kashefi2021}
Kashefi A, Rempe D, Guibas LJ.
\newblock {A point-cloud deep learning framework for prediction of fluid flow
  fields on irregular geometries}.
\newblock Physics of Fluids. 2021;33(2).

\bibitem{Laubscher2021}
Laubscher R.
\newblock {Simulation of multi-species flow and heat transfer using
  physics-informed neural networks Simulation of multi-species flow and heat
  transfer using physics-informed neural networks}.
\newblock Phys Fluids. 2021;33(July).

\bibitem{Pavlova2016}
Pavlova A, Hansen J, Obermeyer H, Pavlova I.
\newblock Geothermal heat in a heat pump use.
\newblock {IOP} Conference Series: Earth and Environmental Science. 2016
  9;43:012025.
\newblock Available from: \url{https://doi.org/10.1088/1755-1315/43/1/012025}.

\bibitem{Kinzelbach1987}
Kinzelbach W.
\newblock Numerische Methoden zur Modellierung des Transports von Schadstoffen
  im Grundwasser.
\newblock 2nd ed. Munich: Oldenbourg; 1987.

\bibitem{pflotrantheory}
PFLOTRAN. PFLOTRAN Documentation: Theory Guide - TH Mode (Thermal-Hydrologic);
  2022.
\newblock Available from:
  \url{https://documentation.pflotran.org/theory\_guide/mode\_th.html}.

\bibitem{Tang2020}
Tang M, Liu Y, Durlofsky LJ.
\newblock A deep-learning-based surrogate model for data assimilation in
  dynamic subsurface flow problems.
\newblock Journal of Computational Physics. 2020;413:109456.

\bibitem{Ronneberger2015}
Ronneberger O, Philipp F, Thomas B.
\newblock U-Net: Convolutional Networks for Biomedical Image Segmentation.
\newblock In: Medical Image Computing and Computer-Assisted Intervention --
  MICCAI 2015. Cham: Springer International Publishing; 2015. p. 234-41.

\bibitem{pflotran-paper}
Hammond GE, Lichtner PC, Mills RT.
\newblock Evaluating the performance of parallel subsurface simulators: An
  illustrative example with PFLOTRAN.
\newblock Water Resources Research. 2014;50:208-28.

\bibitem{Kingma2014}
Kingma DP, Ba J.
\newblock Adam: A Method for Stochastic Optimization. 2014.
\newblock Available from: \url{https://arxiv.org/abs/1412.6980}.

\bibitem{darus-31842022}
Davis K, Schulte M.
\newblock {Replication Data for: Geothermal-ML - predicting thermal plume from
  groundwater heat pumps}. 2022.
\newblock Available from: \url{https://doi.org/10.18419/darus-3184}.

\end{thebibliography}

\end{document}